\documentclass[fleqn,usenatbib]{mnras}

\usepackage{newtxtext,newtxmath}
\usepackage{longtable}

\usepackage[T1]{fontenc}
\usepackage{ae,aecompl}

\usepackage{etoolbox}
\makeatletter
\patchcmd\@combinedblfloats{\box\@outputbox}{\unvbox\@outputbox}{}{\errmessage{\noexpand patch failed}}
\makeatother

\usepackage{graphicx}	
\usepackage{amsmath}	
\usepackage{amssymb}	

\title[Velocities of young radio pulsars II]{The observed velocity distribution of young pulsars II: analysis of complete PSR$\pi$ }

\author[A.P. Igoshev]{
Andrei P. Igoshev,$^{1}$\thanks{E-mail: ignotur@gmail.com}
\\
$^{1}$Department of Applied Mathematics, University of Leeds, Leeds LS2 9JT , UK
}

\date{Accepted XXX. Received YYY; in original form ZZZ}

\pubyear{2015}

\begin{document}
\label{firstpage}
\pagerange{\pageref{firstpage}--\pageref{lastpage}}
\maketitle

\begin{abstract}
Understanding the natal kicks, or birth velocities, of neutron stars are essential for understanding the evolution of massive binaries as well as double neutron star formation.
We use maximum likelihood methods as published in Verbunt et al. to analyse a new large dataset of parallaxes and proper motions measured by Deller et al. This sample is roughly three times larger than number of measurements available before.
For both the complete sample and its younger part (spin-down ages $\tau < 3$~Myr), we find that a bimodal Maxwellian distribution describes the measured parallaxes and proper motions better than a single Maxwellian with probability of 99.3 and 95.0 per cent respectively. The bimodal Maxwellian distribution has three parameters: fraction of low-velocity pulsars and distribution parameters $\sigma_1$ and $\sigma_2$ for low and high-velocity modes.
For a complete sample, these parameters are as follows: $42_{-15}^{+17}$ per cent, $\sigma_1=128_{-18}^{+22}$~km~s$^{-1}$ and $\sigma_2 = 298\pm 28$~km~s$^{-1}$. For younger pulsars, which are assumed to represent the natal kick,  
these parameters are as follows: $20_{-10}^{+11}$~per~cent, $\sigma_1=56_{-15}^{+25}$~km~s$^{-1}$ and $\sigma_2=336\pm 45$~km~s$^{-1}$. In the young population, $5\pm 3$ per cent of pulsars has velocities less than 60~km~s$^{-1}$.
We perform multiple Monte Carlo tests for the method taking into account realistic observational selection. We find that the method reliably estimates all parameters of the natal kick distribution. Results of the velocity analysis are weakly sensitive to the exact values of scale-lengths of the Galactic pulsar distribution.

\end{abstract}




\section{Introduction}

The origin of pulsar velocities is of profound importance to many areas of astrophysics. Of particular interest is the distribution of \textit{natal kicks}.
Natal kicks are defined here as the velocity received by neutron star (NS) in excess to its progenitor local standard of rest (LSR) velocity (if formed from isolated massive star), or as the velocity received by the NS, which disrupted a binary in excess to the LSR velocity of the binary centre of mass. It was shown by \cite{kuranov2009} that even in the case of binary origin the observed space velocities are mostly affected by the natal kick velocity distribution and not by binary evolution. 
NSs are known to receive natal kicks at the moment of supernova explosion \citep{evidence_nskick}.
This fact is derived from observations of large peculiar velocities (order of 100~km~s$^{-1}$) of isolated radio pulsars in comparison to their progenitors O-B stars with peculiar velocities of $\approx 10-15$~km~s$^{-1}$ \citep{lynelorimer}. The young NSs are known to be located at some offset from the centre of the associated supernova remnants (SNRs; \citealt{SNRremnantoffset,NS_SNR_new}). Some NSs demonstrate a bow shock as a proof of direct association of the pulsar with the SNR and large pulsar speed e.g. the Guitar nebula \citep{guitar_nebula}.
The scale height of radio pulsar population in the Galaxy ($330$~pc; \citealt{lorimer2006}) is much larger than the scale height of pulsar progenitors ($\approx 50$~pc). These multiple direct observations 
confirm a large natal kick scenario for a significant number of NSs.

There are certain indications that some NSs received very small natal kicks. In particular, the formation of some double NSs systems requires natal kicks of order of tens km~s$^{-1}$ \citep{tauris_dns}. \cite{lowkickNSNS} used binary population synthesis coupled together with cosmological simulation to study dependence of double NSs merger rate on natal kicks. They found that it was necessary to assume extremely small value of $\sigma=15$~km~s$^{-1}$ for the Maxwellian velocity distribution to reproduce the double NSs merger rate derived from the gravitational wave detection GW170817 \citep{double_NSmerger}. There is a group of Be X-ray binaries with small eccentricities and large orbital periods \citep{pfahl2002,townsed_bex_smc} which require natal kicks of $v < 50$~km~s$^{-1}$ to explain their observational properties. This indirect evidence suggests that a noticeable fraction of NSs are born with small natal kicks. 

Knowledge of the natal kick distribution for neutron stars (NSs) is important because it is an essential ingredient of models for double NSs formation, in particular the double NS mergers \citep{double_NSmerger}, short gamma-ray bursts, millisecond  pulsars formation and different scenarios for white dwarfs-NS mergers \citep{toonen_NSWD}. Natal kicks are essential to model the Galactic distribution of pulsars and eventually design radio surveys to discover new NSs.  A fraction of low-velocity NSs is important to predict and interpret (if discovered) a fraction of NSs accreting from the interstellar medium \citep{accreting_from_ism, Shvartsman1971}. For example, \cite{popov2000} derived the lower bound on the mean kick velocity as 200-300~km~s$^{-1}$ based on absence of isolated accreted NSs sources in the ROSAT observations. Future observations with the eROSITA mission \citep{erosita}, which is much more sensitive than ROSAT and should discover multiple isolated accreted NSs will help solving this issue \citep{erositaPopov}.

One of the well-known ideas to constrain the natal kick velocity distribution is to analyse the parallaxes and proper motions of young isolated radio pulsars, see e.g. \cite*{arzoumanian} and \cite*{verbunt2017}. \cite{verbunt2017} analysed a small sample containing 28 pulsars (19 pulsars with spin-down age less than 10~Myr). They used a maximum likelihood method to estimate parameters of the velocity distribution consistently.  
This estimate crucially depends on our knowledge of precise parallaxes and proper motions for a large number of objects.

The primary reason for this new study is a recent publication of a large sample of  precise interferometric measurements by \cite{Deller2019}.  This sample contains measurements for 57 radio pulsars; most parallaxes are measured for the first time. We use the same maximum likelihood technique described by \cite{verbunt2017} to analyse this extended sample combined with older measurements. Therefore, we aim at obtaining a more reliable estimate for the natal kick velocity distribution. 
We also aim at testing the sensitivity of the method to a particular choice of parameters for Galactic distribution of pulsars and observational selection.

The article is structured as follows: in Section 2 we describe our dataset, in Section 3 we briefly introduce the essential ingredients of the maximum likelihood technique and in Section 4 we present the results. We conclude our paper with discussion and conclusions. 





\section{Data}
In this research, we combine new measurements presented by \cite{Deller2019} with older reliable VLBI measurements of parallaxes and proper motions previously collected by \cite{verbunt2017}. Table~\ref{t:lit_sourc} shows the list of literature sources; all the measurements are compiled in the master list Table~\ref{t:master}. To find periods and period derivative of pulsars we use the ATNF catalogue \citep{atnf}\footnote{http://www.atnf.csiro.au/research/pulsar/psrcat}.
In this research we decide to keep only reliable measurements for parallax i.e. $\varpi'/\sigma_\varpi > 3$ where $\varpi'$ is the measured parallax and $\sigma_\varpi$ is the parallax uncertainty. A cut-off $\varpi'/\sigma_\varpi > 2$ or $\varpi'/\sigma_\varpi > 4$ removes just one additional pulsar from the list.  
In principle, our method could deal with upper limits as well, but in this case the final result becomes much more sensitive to exact assumptions about the spatial distribution of the pulsars in the Galaxy. This distribution is based on previous distance measurements (often via the electron density model) and is not particularly reliable for radio pulsars. If the parallax or proper motion uncertainty is non-symmetric, we choose the greatest value out of two.

 To update the list we replace parallax and proper motion measurements by \cite{bbgt02} with values provided by \cite{Deller2019} for two radio pulsars: J0332+5434 (new parallax differs $3.05\sigma$) and J1136+1551 (new parallax differs $0.71\sigma$). \cite{Deller2019} provided parallax measurements for five pulsars from \cite{bfg+03} with measured proper motions. Three of theses pulsars (J1645-0317, J1735-0724 and J2305+3100) are younger than spin-down age $\tau = 10$~Myr. As a measure of the pulsar age we use the spin-down age $\tau = P / (2\dot P)$ where $P$ is the pulsar period and $\dot P$ is the period derivative. The spin-down age could differ from real age if the initial period of pulsar is large or the pulsar experienced a magnetic field evolution. It seems that the spin-down age works reasonably well for most pulsars with $\tau < 10-20$~Myr \citep{igoshev_age2019}.

In general our master list includes nearly all objects from \cite{Deller2019} with exception of four millisecond (recycled) radio pulsars (J2010-1323, J2145-0750, J2317+1439 and J1022+1001). We  also exclude NSWD system J0823+0159. Binaries with radio pulsars move in respect to the local standard of rest with a peculiar velocity which could significantly differ from the pulsar natal kick because the energy and momentum of the kick are partly redistributed into the orbital parameters.
As in the previous research we do not include any pulsars located inside globular clusters.
The total number of previous measurements of good quality is 24, while the number of new good quality measurements is 45. We do not extend our sample further to include pulsars with distances based on dispersion measure and an electron density model of the Galaxy as it was shown by \cite{Deller2019} these distances could differ 3-5 times from parallax measurements both for NE2001 \citep{ne2001} and YMW16 \citep{ymw16} models. 

\begin{table}
    \centering
    \begin{tabular}{clcccc}
    \hline
    S & Source            &          & N & n \\    
    \hline
    1 & \cite{bbgt02}     & Table 4  & 4 & 0 \\
    2 & \cite{btgg03}     & Table 3  & 1 & 1 \\
    3 & \cite{ccl+01}     & Table 2  & 1 & 1 \\
    4 & \cite{ccv+04}     & Table 1  & 1 & 1 \\
    5 & \cite{dtbr09}     & Table 3  & 3 & 1 \\
    6 & \cite{cbv+09}     & Table 2  & 11 & 5 \\
    7 & \cite{kvw+15}     & Table 5  & 3  & 2 \\
    8 & \cite{Deller2019} & Table 3  & 45 & 10 \\
      &                   & Total:   &  69  & 21 \\ 
    \hline  
    \end{tabular}
    \caption{Sources for parallaxes and proper motions in our master list. N is a number of objects with $\varpi'/\sigma_\varpi > 3$ and $n$ is a number of those younger than $3$~Myr.}
    \label{t:lit_sourc}
\end{table}

\begin{figure*}
    \centering
    \includegraphics[width=1.0\linewidth]{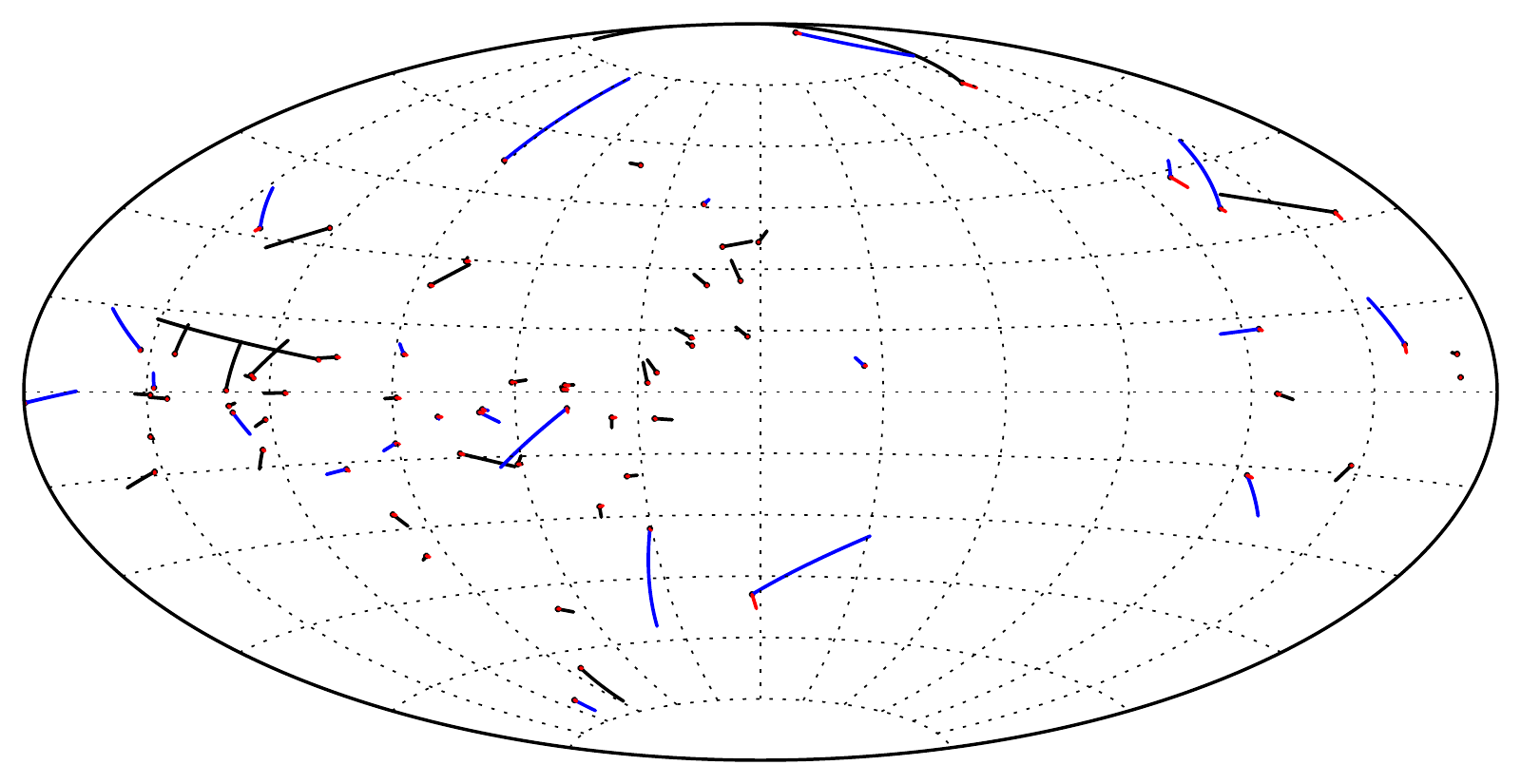}
    \caption{Proper motions of pulsars in 0.5 Myr (blue lines for older measurements, black lines for objects form \protect\citealt{Deller2019}) in the Galactic coordinate system. Red lines show the proper motion  with respect to the local standard of rest at each position (see text). }
    \label{f:map}
\end{figure*}
\begin{figure*}
    \centering
    \begin{minipage}{0.48\linewidth}
    \includegraphics[width=1.0\linewidth]{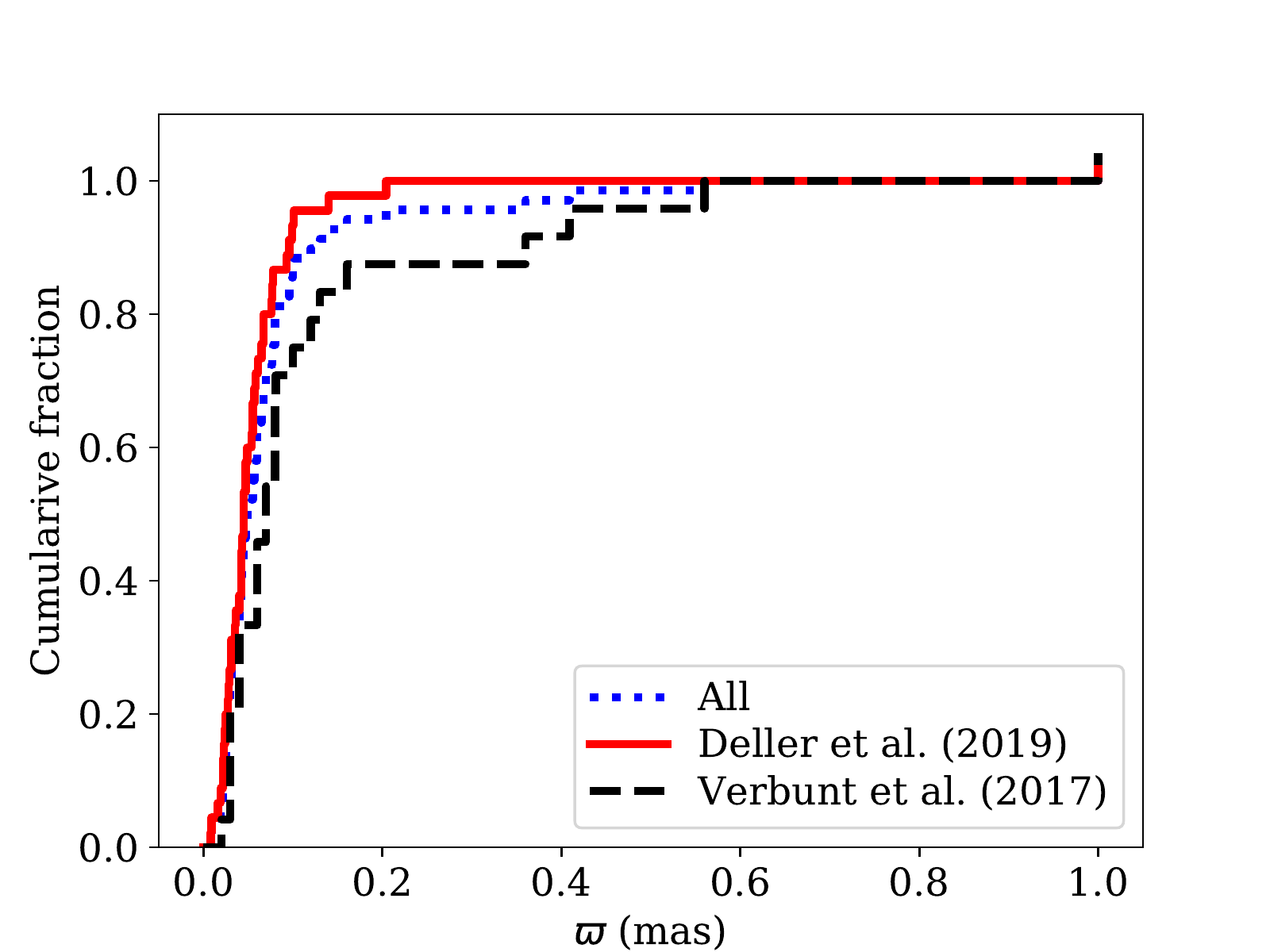}
    \end{minipage}
    \begin{minipage}{0.48\linewidth}
    \includegraphics[width=1.0\linewidth]{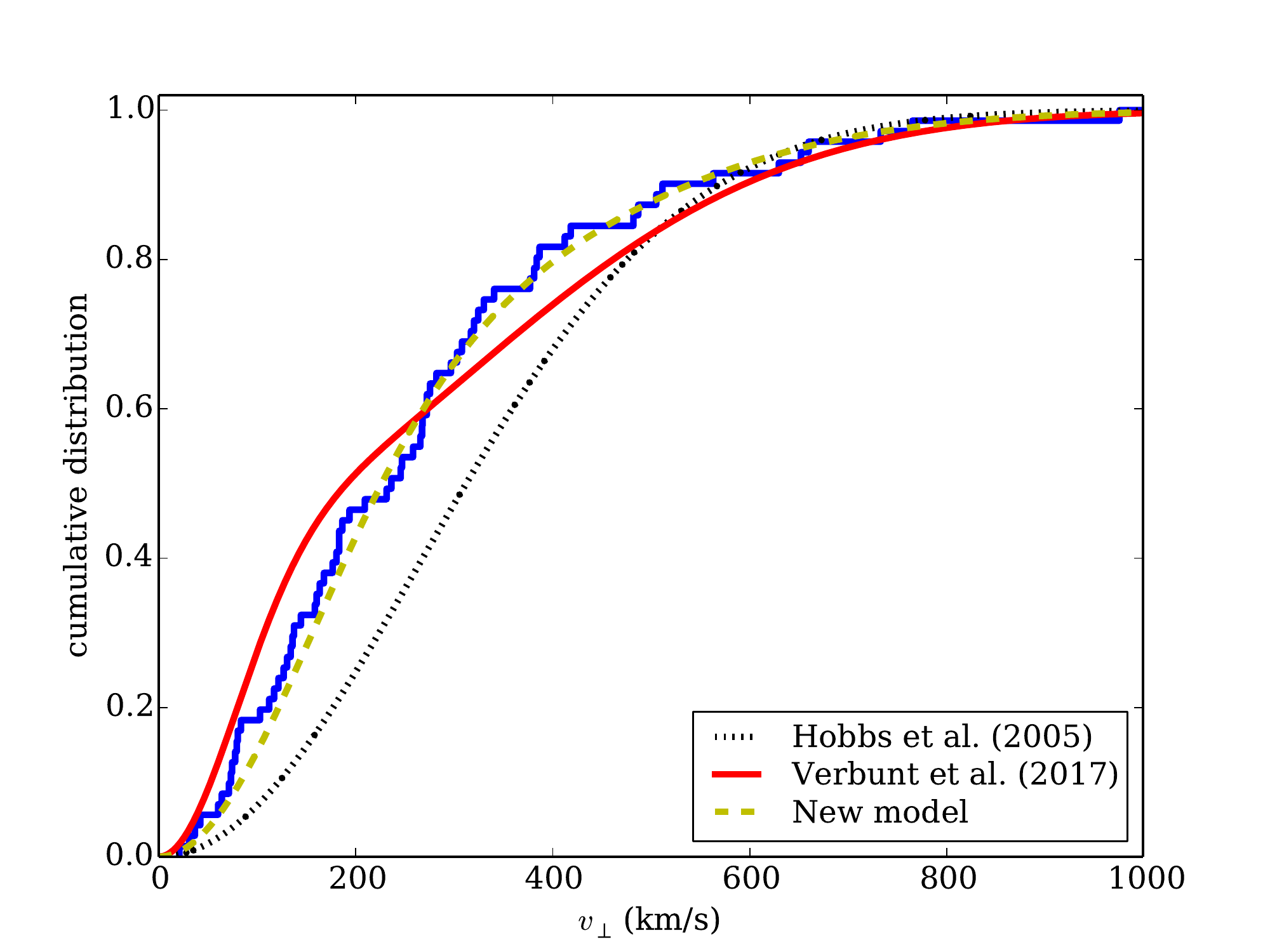}
    \end{minipage}
    \caption{Left panel: cumulative distribution of measured parallaxes for pulsars from \protect\cite{verbunt2017} (dashed black line), from \protect\cite{Deller2019} (solid red line) and combination of these two catalogues (dotted blue line). Right panel: the cumulative distribution of transverse velocities for radio pulsars with well measured parallaxes ($\varpi/\sigma_\varpi>3$; blue histogram) and cumulative velocity distributions presented earlier in \protect\cite{hobbs2005} (dotted black line) and \protect\cite{verbunt2017} (solid red line). Dashed yellow line show result of optimisation for all pulsars using the mixed model.}
    \label{f:v_trans_cumul}
\end{figure*}

We plot the positions and proper motions for radio pulsars in our master list in Figure~\ref{f:map}. 
It is easy to notice a dearth of objects around $l=270$ (possibly because it is unavailable for the VLBI observations). In general we notice objects above and below the Galactic plane. 
Despite the corrections for motion of LSR being small as it is seen from Figure~\ref{f:map} (red lines), we take them into account by adopting a flat Galaxy rotation curve with speed $v_R = 220$~km~s$^{-1}$ and distance of the Sun from the Galactic center $R_\odot = 8.5$~kpc. 
Pulsars in the catalogue by \cite{Deller2019} have smaller parallaxes in comparison to previous measurements, see left panel of Figure~\ref{f:v_trans_cumul}.

In right panel of Figure~\ref{f:v_trans_cumul}, we plot the cumulative distribution of transverse velocities computed as:
\begin{equation}
v_\perp [\mathrm{km~s}^{-1}] = \frac{4.74}{ \varpi'[\mathrm{mas}]} \sqrt{{\mu'}_{\alpha *}^2 + {\mu'}_\delta^2}\,\,[\mathrm{mas~yr}^{-1}] 
\end{equation}
where ${\mu'}_{\alpha *}$ and ${\mu'}_\delta$ are the measured values of proper motion in the direction of right ascension and declination. We see that the distribution suggested by \cite{hobbs2005} has a systematic shift, and therefore does not describe the new data. The origin of this systematic shift could be due to overestimated distances to pulsars.
The distribution suggested in \cite{verbunt2017} seems to reliably estimate the low-velocity tail, but somewhat differs from the data in regions around $\approx 100$~km~s$^{-1}$ and $\approx 400$~km~s$^{-1}$.

\section{Method}
We use the same maximum likelihood methods as described in \cite{verbunt_conf} and \cite{verbunt2017}, therefore we refer any interested reader to these two articles for detailed description. Here we give a brief summary highlighting most relevant details.

\subsection{The Galactocentric pulsar distribution}

We assume that the Galactic distribution of pulsars forms a thin, exponential disk. In this assumption we follow the previous works of \cite{verbiest2012} and \cite{lorimer2006} with corrections of \cite{igoshev2016}. The exact form of the distribution is:
\begin{equation}
f_D(D) dD = C D^2 R^{1.9} \exp \left[-\frac{|z|}{h}-\frac{R}{H}\right] dD,   
\label{e:fD}
\end{equation}
where $|z|$ is the height of a pulsar above the Galactic plane, $C$ is the normalisation factor, $R$ is the Galactocentric distance of the pulsar. Two values $h$ and $H$ are parameters of the model; for young radio pulsars these values are usually assumed to be $h=0.33$~kpc and $H=1.7$~kpc. Given the wealth of interferometric data provided by \cite{Deller2019}, we aim to optimise these parameters. Eq. (\ref{e:fD}) is normalised numerically for each direction and each set of parameters $h, H$ for $D \in [0, 10]$~kpc, such that:
\begin{equation}
\int_0^{10} f_D (D)\, dD = 1.    
\end{equation}
This step is essential if parameters $h$ and $H$ are optimised using the maximum likelihood technique.

\subsection{Velocity distributions}

We analyse four different velocity distributions $f_v (\vec v | \vec \sigma)dv$ with vector of parameters $\vec \sigma$. First, we use isotropic Maximilian velocity distribution\footnote{The Maxwellian velocity distribution is isotropic by definition, but because we want to study a modified Maxwellian distribution made anisotropic, we highlight this fact} with a single parameter $\sigma$:
\begin{equation}
f_{v, \mathrm{maxw}} (\vec v | \sigma) dv = \sqrt{\frac{2}{\pi}} \frac{v^2}{\sigma^3} \exp\left[ -\frac{v^2}{2\sigma^2} \right] dv; \hspace{1cm} (0<v<\infty)  \label{e:maxw}   
\end{equation}
Second, we use a sum of two isotropic Maxwellian distributions (the so-called bimodal Maxwellian distribution):
$$
f_{v, 2\mathrm{maxw}} (\vec v | w, \sigma_1, \sigma_2)\, dv = w f_{v, \mathrm{maxw}} (\vec v | \sigma_1)\hspace{2.5cm}
$$
\begin{equation}
\hspace{4cm}+ (1 - w) f_{v, \mathrm{maxw}} (\vec v | \sigma_2)\, dv,
\label{e:2maxw}
\end{equation}
where $w$ is a fractional contribution of the first Maxwellian, such that $w\in [0, 1]$. If $w = 0$ or $w=1$ the distribution defined by eq. (\ref{e:2maxw}) becomes a single Maxwellian. 

Third, we introduce a mixed model which is a combination of isotropic and a semi-isotropic Maxwellian distribution $f_{v, \mathrm{semi}} (\vec v | \sigma) dv$. The latter one is a Gaussian distribution for velocity component directed away from the Galactic plane ($v_z \cdot z > 0$) and zero otherwise, two other velocity components are not affected. 
The motivation for this model is as follows: young pulsars born in the Galactic plane with velocities pointing outside the plane move for some time (30-100 Myr) in the same direction as their original velocity vector. Meaning that if a young pulsar is discovered at positive galactic latitude $b>0$, most probably its $\mu_b$ is also positive. An older pulsar could also move toward the plane because its velocity was altered significantly by the Galactic gravitational potential. Therefore we cannot apply the semi-isotropic model to them. This model is a good first test for probing the natal kick velocity orientation in the younger sample.
If the semi-isotropic model provides a better fit than the isotropic model, it means that their speeds were not significantly altered by Galactic gravity. The model is:
\begin{equation}
f_{v, \mathrm{semi}} (\vec v | \sigma) dv  =
\left\{
\begin{array}{lc}
2 \sqrt{\frac{2}{\pi}} \frac{v^2}{\sigma^3} \exp\left[ -\frac{v^2}{2\sigma^2} \right] dv; & \mathrm{if}\; v_z \cdot z > 0  \\
0; & \mathrm{otherwise}
\end{array}
\right.
\end{equation}
We call it a mixed velocity distribution because we apply the isotropic Maxwellian distribution to one group of pulsars and semi-isotropic Maxwellian distribution to another group. 
The first group (isotropic) contains 17 pulsars with no clear preference for orientation of the velocity vector. The second group (semi-isotropic) contains 52 pulsars with expected motion directed away from the Galactic plane.
The second group consists of young and intermediate age pulsars (spin-down age $\tau < 50$~Myr with nominal height $|z| = |\sin b / \varpi'| > 50$~pc), first group includes all remaining pulsars. 

Fourth, we combine two mixed Maxwellian distributions:
$$
f_{v, 2\mathrm{mix}} (\vec v | w, \sigma_1, \sigma_2)\, dv = w f_{v, \mathrm{mix}} (\vec v | \sigma_1)\hspace{2.5cm}
$$
\begin{equation}
\hspace{4.5cm}+ (1 - w) f_{v, \mathrm{mix}} (\vec v | \sigma_2)\,  dv.
\label{e:2maxw_mix}
\end{equation}
This model is analogous to model which consists of two isotropic Maxwellains. It is introduced here to test if the sample is indeed kinematically young.

\subsection{The likelihood function and model comparison}
Using eq. (\ref{e:fD}) and one of velocity distributions eqs. (\ref{e:maxw}-\ref{e:2maxw_mix}) together with normal distributions for measurement errors for parallax $g_\varpi (\varpi' | D)$ and two components of proper motion $g_\mu (\mu_{\alpha*}' | \vec v, D)$, $g_\mu (\mu_{\delta}' | \vec v, D)$, we compute the joint probability $P (\varpi', \mu_{\alpha*}' \mu_{\delta}', \vec v,D | \vec \sigma)$. 
The joint probability depends on unknown actual distance $D$ and velocity vector $\vec v$, so we integrate it over these quantities partly numerically and partly analytically to produce the conditional probability of individual measurements given a parameter of the particular velocity distribution:
\begin{equation}
P(\varpi', \mu_{\alpha*}' \mu_{\delta}'| \vec \sigma) = \iiiint    P (\varpi', \mu_{\alpha*}' \mu_{\delta}', \vec v,D | \vec \sigma)\, d^3 \vec v \, dD.
\end{equation}
We assume that the measurements of individual pulsars do not depend of each other, in this case the conditional probabilities for different pulsars can be multiplied to form a likelihood of the model given the data. Doing so, we also assume that the proper motion measurement is independent of the parallax measurement. Measurements of proper motion might correlate with parallax measurements but the value of this correlation is not currently provided in the datasets. 
For numerical convenience we compute the logarithm of the likelihood:
\begin{equation}
\mathcal L (\vec \sigma) = -2\sum_{i=0}^{N} \log P(\varpi', \mu_{\alpha*}' \mu_{\delta}'| \vec \sigma).
\end{equation}
We further optimise the log-likelihood for every velocity distribution and find the best parameters $\hat \sigma$.

We use the log-likelihood difference:
\begin{equation}
\Delta \mathcal {L} = \mathcal L (\sigma) - \mathcal L(\hat\sigma),    
\end{equation}
in each direction of the parameter vector $\vec \sigma$ to estimate the confidence interval. The log-likelihood difference of $\Delta \mathcal L = 1$ approximately corresponds to a 68 per cent confidence region. 

\begin{figure}
    \centering
    \includegraphics[width=1.0\linewidth]{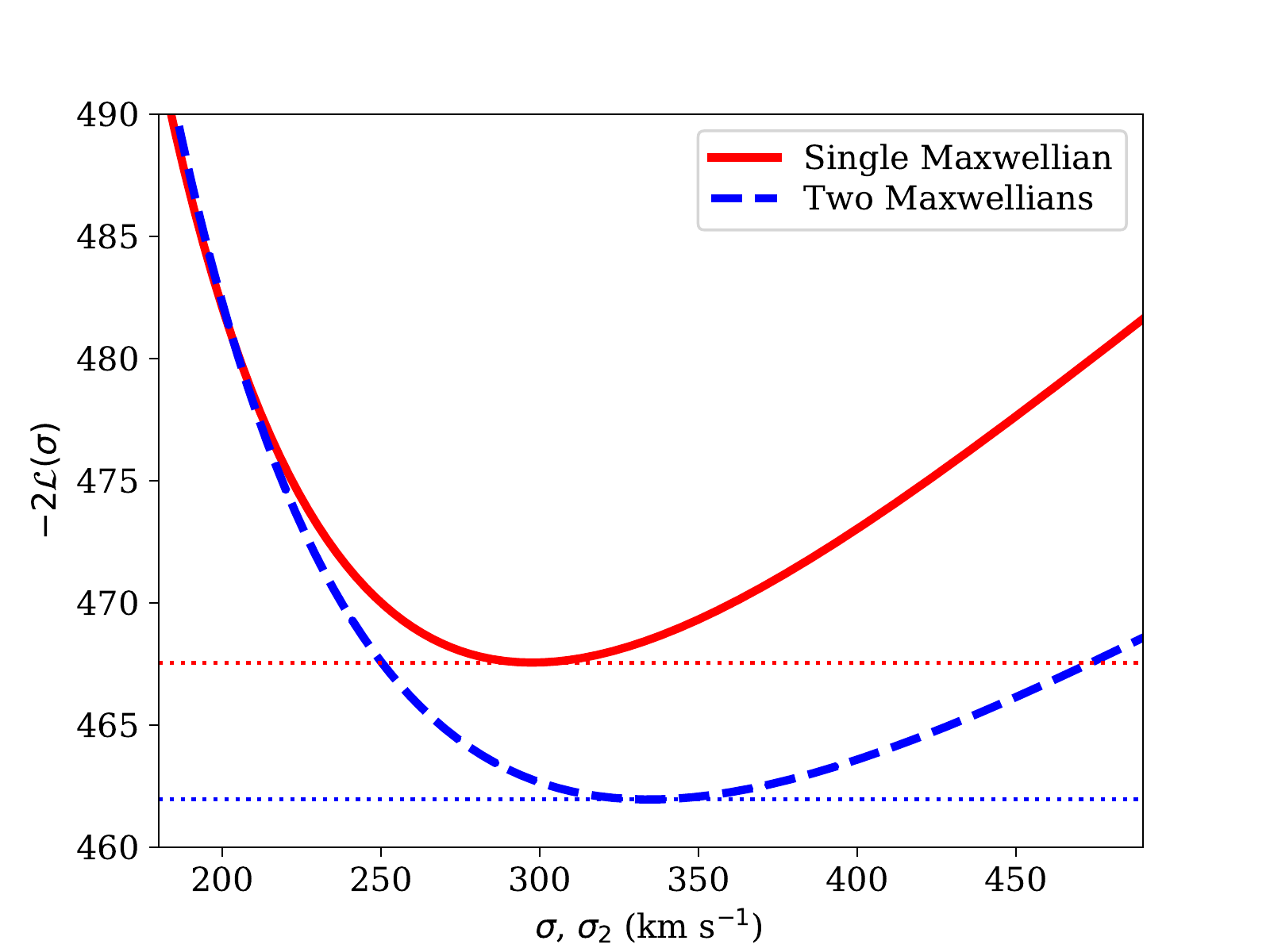}
    \caption{The log-likelihood profile for Y sample for model with a single isotropic Maxwellian distribution (red solid curve) and a sum of two isotropic Maxwellian distributions (blue dashed line). In the case of the latter model we fix $w=0.19$ and $\sigma_1 = 55$~km~s$^{-1}$. The minima are indicated by horizontal dotted lines for each case.}
    \label{f:like_prof}
\end{figure}

To compare two velocity distributions given the same dataset (with respective log-likelihoods $\mathcal L^a (\hat \sigma_1)$ and $\mathcal L^b (\hat \sigma_2)$), we compute log-likelihood difference of these models:
\begin{equation}
d\mathcal L = \mathcal L^a (\hat \sigma_1) - \mathcal L^b (\hat \sigma_2).
\end{equation}
This is the likelihood ratio test, which follows the asymptotic of $\chi^2$ distribution with number of parameters equal to number of additional free variables present in the second model in comparison to the first model. We illustrate the model comparison in Figure~\ref{f:like_prof}. The likelihood function for a model with a single Maxwellian reaches maximum around 300~km~s$^{-1}$ with a value of log-likelihood $2\mathcal{L} = -467$. On the other hand the likelihood function for sum of two isotropic Maxwellians reaches higher maximum at $\sigma_2 \approx 340$~km~s$^{-1}$ with $2\mathcal {L} = -461$. The likelihood difference $d\mathcal L = 6$. Thus the model with a single Maxwellian distribution is rejected at the level of 5~per cent because $\chi^2 = 6$ corresponds to 95 per cent probability for two degrees of freedom.    The usage of the likelihood ratio test is justified for comparison of models consistent of two Maxwellians versus a single Maxwellian since these are nested models i.e. two Maxwellians can approximate a single Maxwellian distribution if parameters are chosen in a particular way. We confirm that the likelihood ratio test works for model comparison using Monte Carlo tests described in Appendix~\ref{s:tests}.

\section{Results}
\subsection{Velocity distribution of all radio pulsars}
We show the results of our analysis in Table~\ref{t:res}.
Overall,  a Maxwellian with $\sigma=229_{-14}^{+16}$~km~s$^{-1}$ fits the distribution. \cite{verbunt2017} give an estimate of $\sigma=244$~km~s$^{-1}$ which is within the 68~per~cent confidence interval. In comparison to that work, the confidence interval shrunk from $\approx 50$~km~s$^{-1}$ to $\approx 30$~km~s$^{-1}$ as the number of pulsars in the sample increased. The result of \cite{hobbs2005}  ($\sigma=265$~km~s$^{-1}$) is outside of 68 per cent confidence interval, but is within 99 per cent confidence interval.

\begin{figure}
    \centering
    \includegraphics[width=1.0\linewidth]{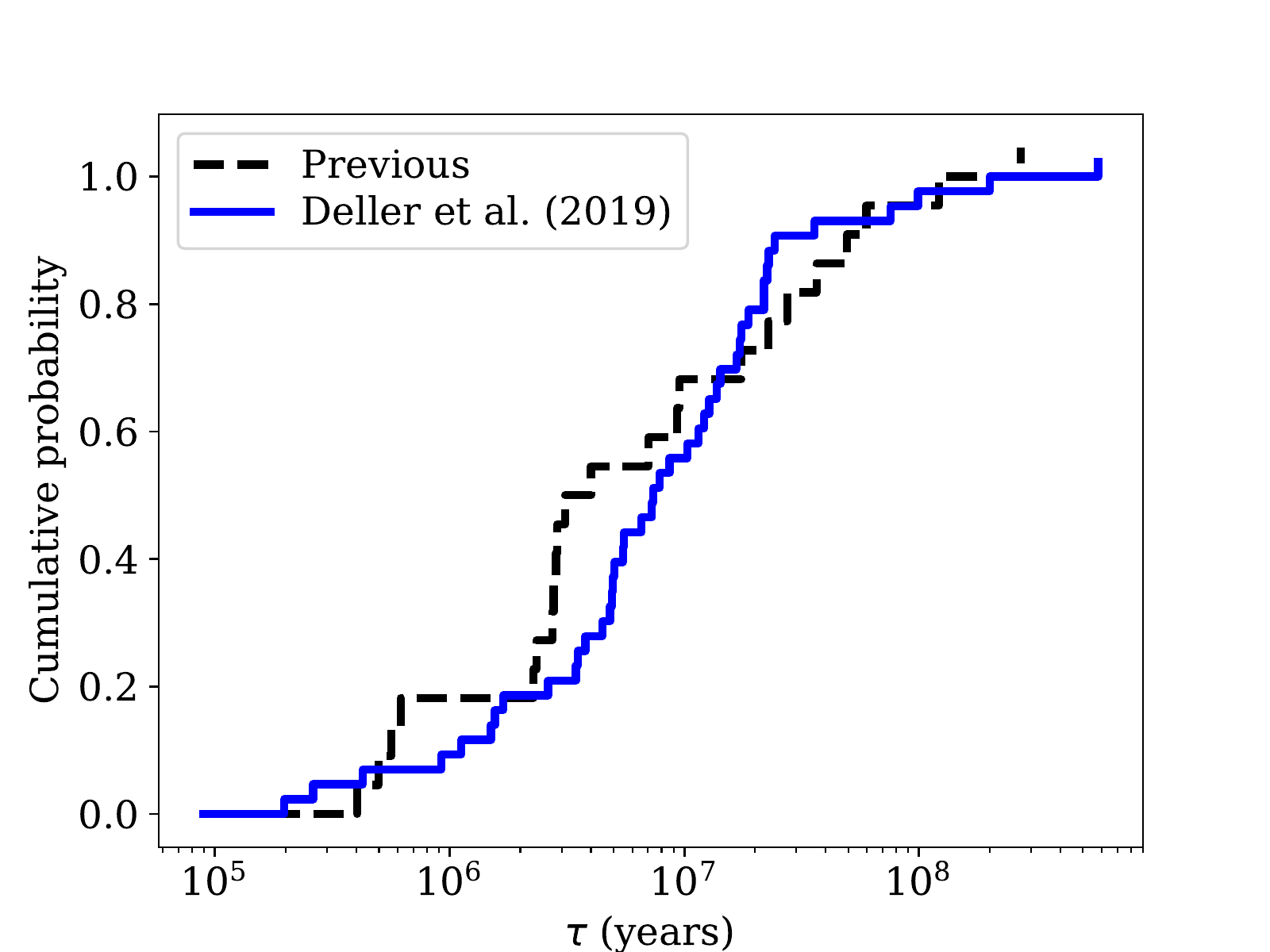}
    \caption{Cumulative distribution of spin-down ages for pulsars in sample of \protect\cite{verbunt2017} (black dashed line) and in the sample of \protect\cite{Deller2019} (solid blue line).}
    \label{f:tau}
\end{figure}

\begin{figure*}
    \centering
    \begin{minipage}{0.48\linewidth}
    \includegraphics[width=1.0\linewidth]{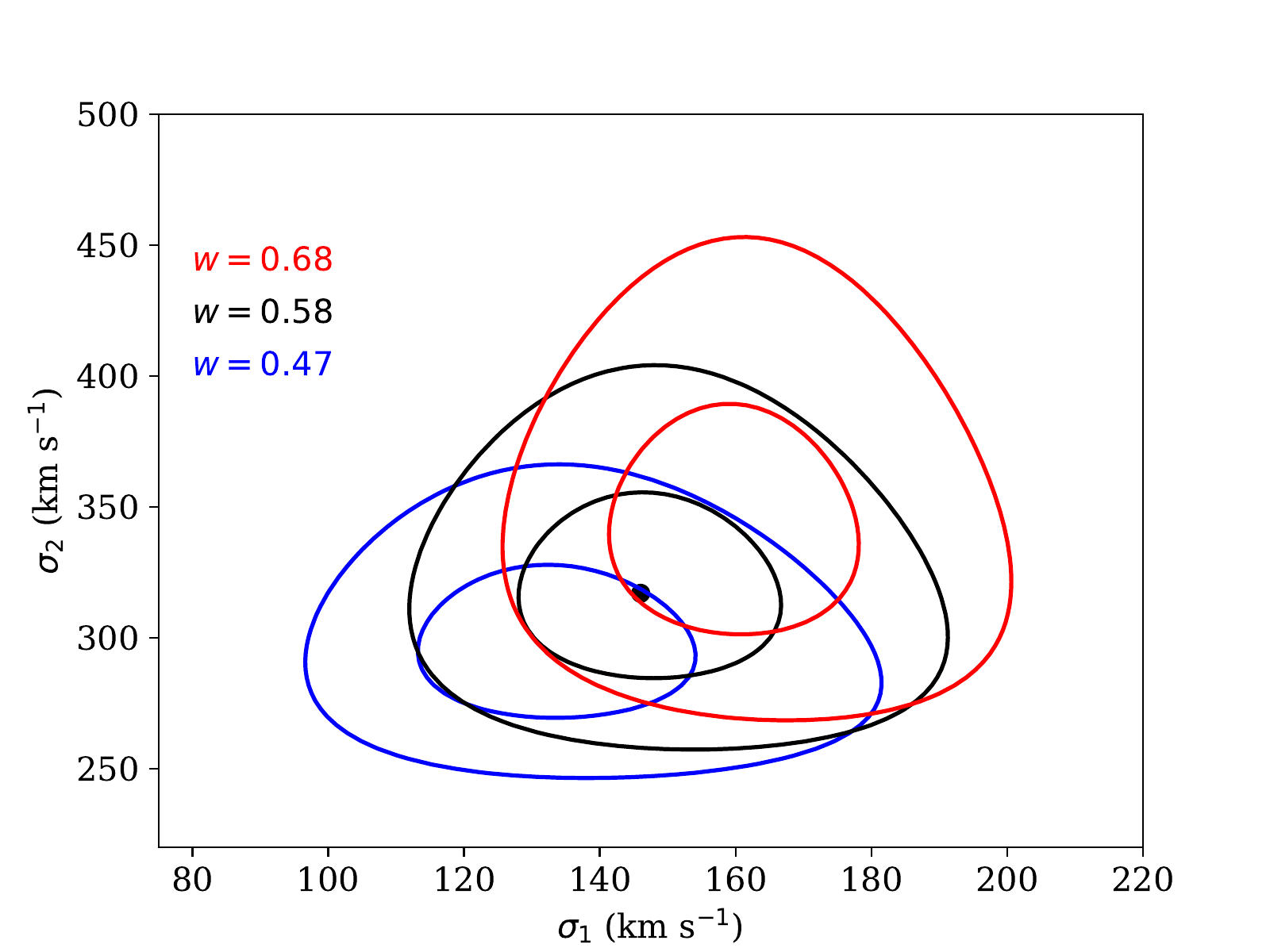}
    \end{minipage}
        \begin{minipage}{0.48\linewidth}
    \includegraphics[width=1.0\linewidth]{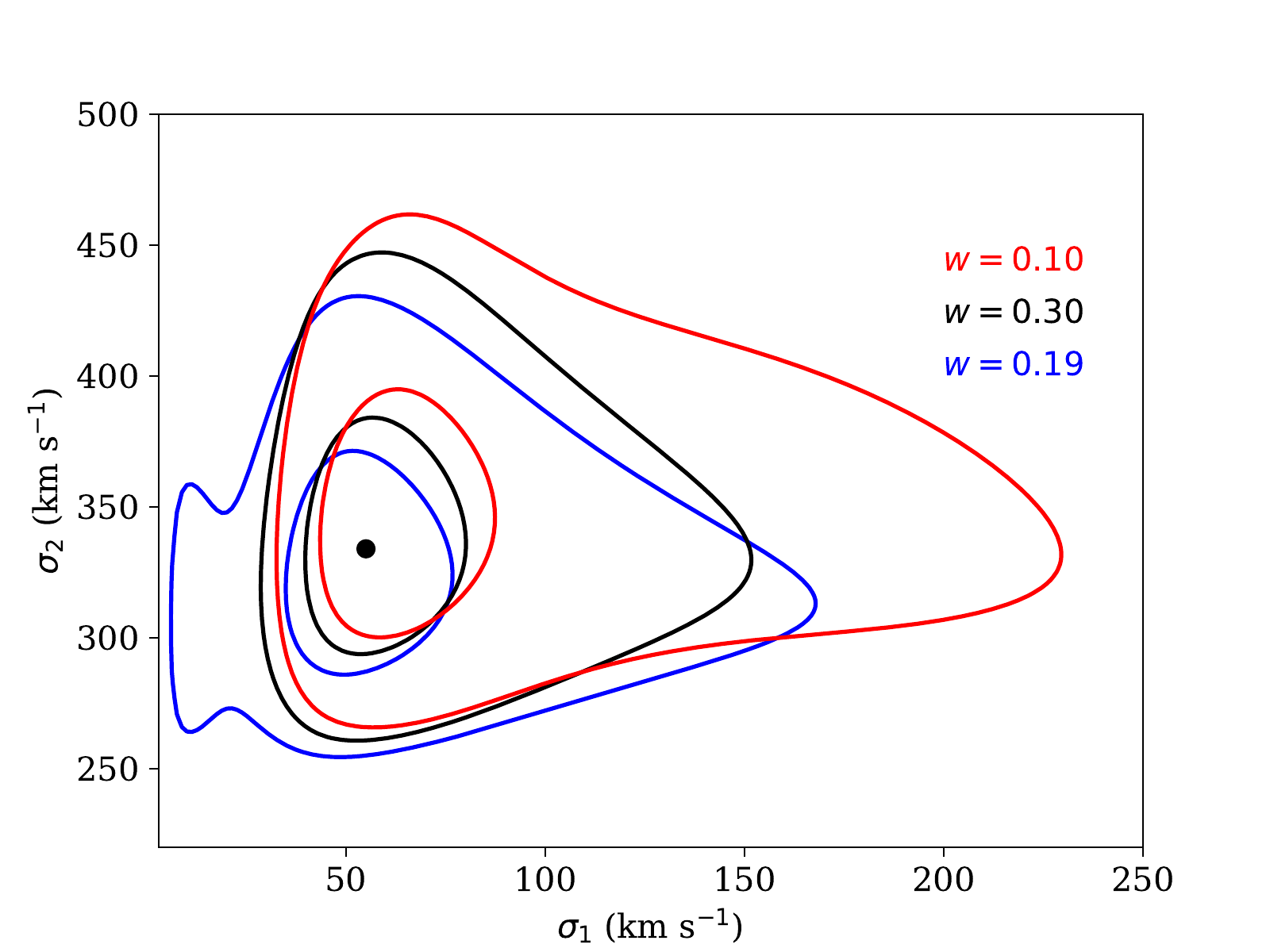}
    \end{minipage}
    \begin{minipage}{0.48\linewidth}
    \includegraphics[width=1.0\linewidth]{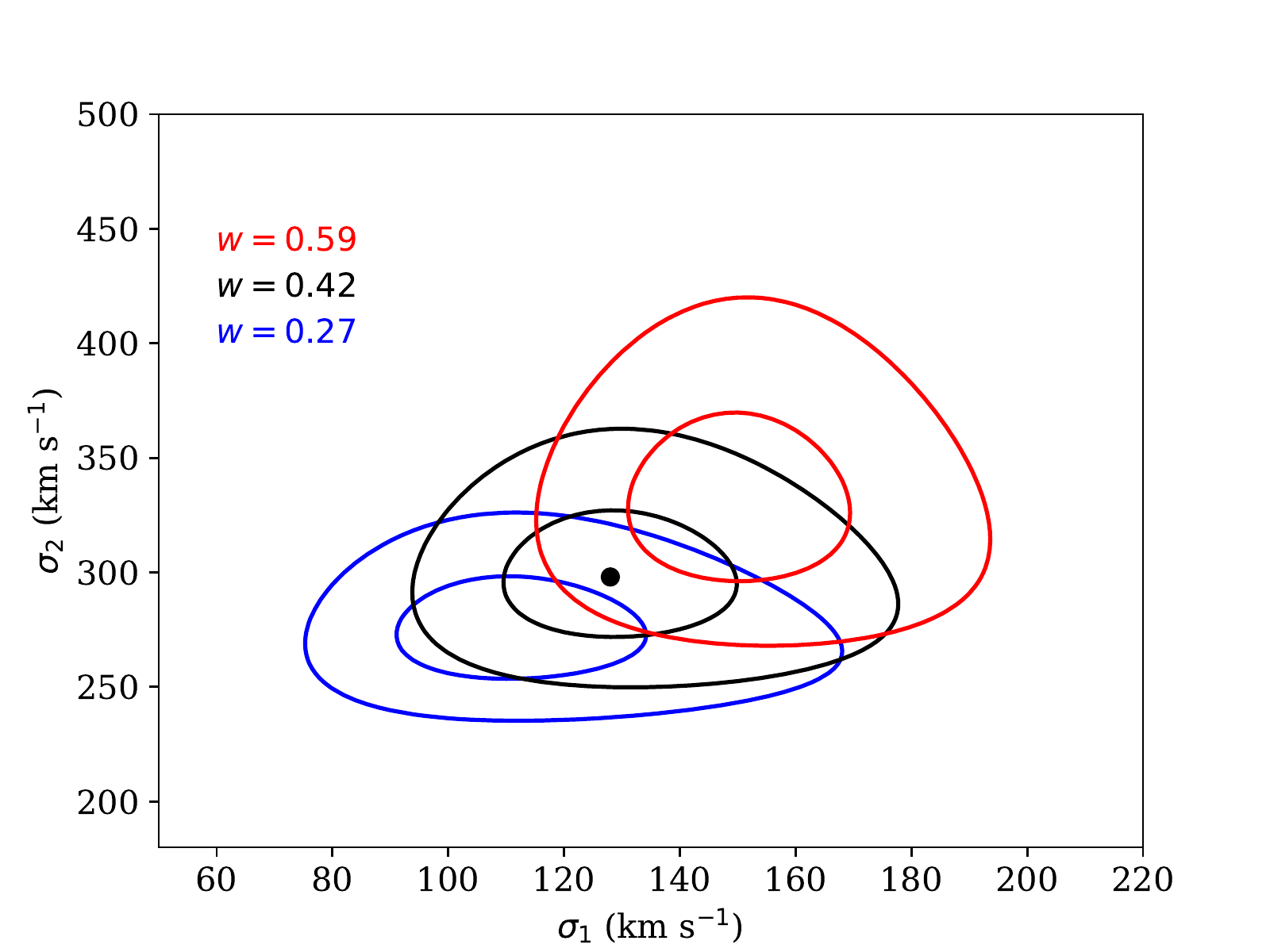}
    \end{minipage}
    \begin{minipage}{0.48\linewidth}
    \includegraphics[width=1.0\linewidth]{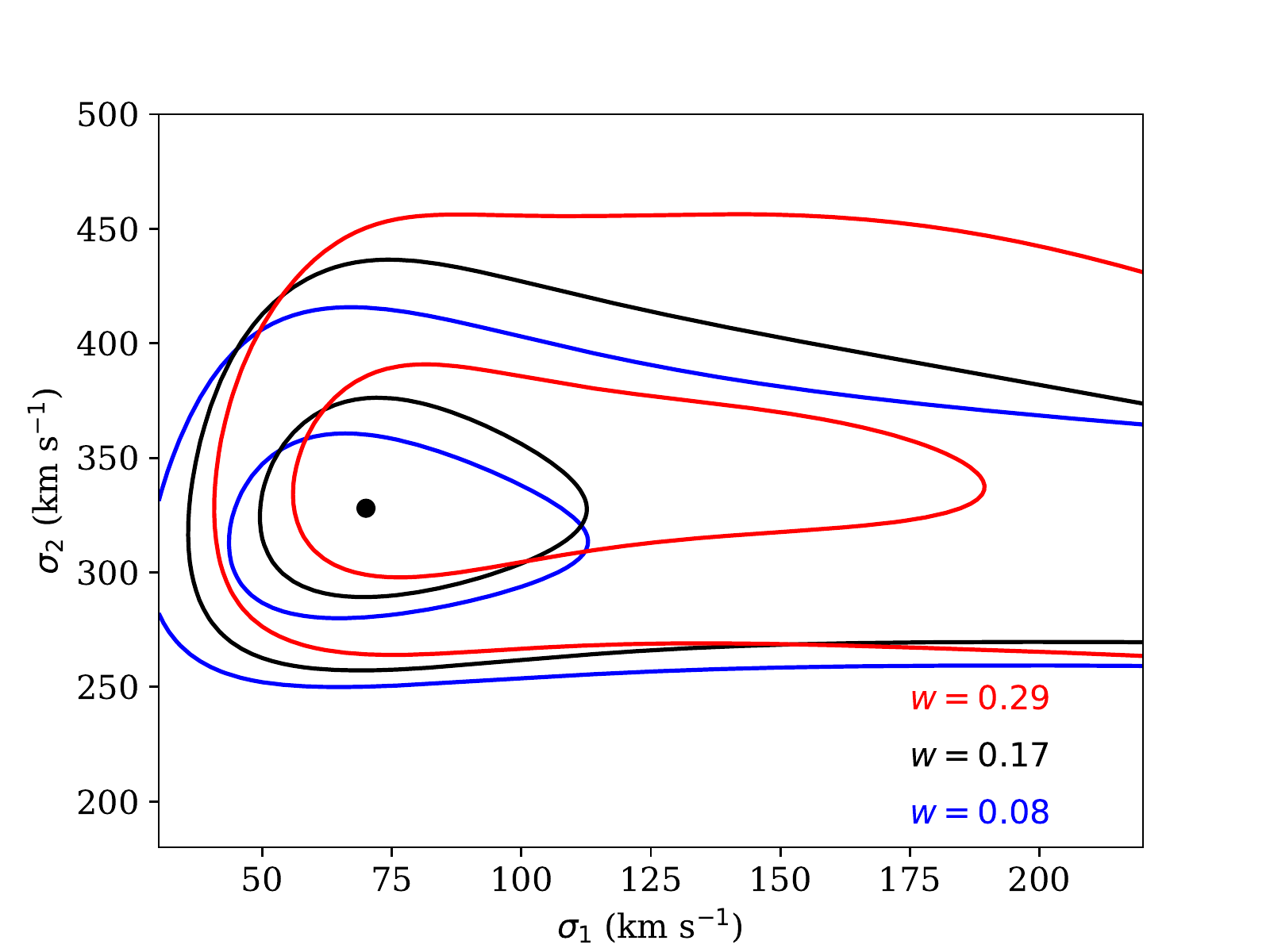}
    \end{minipage}
     \caption{Contour of $d\mathcal L(\sigma)$ for fixed values of $w$ for all pulsars in our sample (left column) and only young radio pulsars ($\tau < 3$~Myr; right column). In the upper row we show results for isotropic model and in lower row we show results for the mixed model.}
    \label{f:contour_A}
\end{figure*}

When we use the isotropic bimodal Maxwellin distribution we get following result: $58_{-11}^{+10}$~per cent of objects are drawn from the Maxwellian distribution with $\sigma_1 = 146_{-21}^{+16}$~km~s$^{-1}$ and $42_{-10}^{+11}$~per~cent from the Maxwellian with $\sigma_2=317_{-40}^{+29}$~km~s$^{-1}$.
The contours of constant likelihood are shown in Figure~\ref{f:contour_A} (upper left panel). The most probable $\sigma_1$ correlates with the contribution of the first component $w$: the lower $w$, the lower $\sigma_1$ should be. It could indicate that the actual distribution is not a sum of two Maxwellians, but instead it could be described by two separate variables.  
While the contribution and location of the high-velocity component are roughly consistent with the result of \cite{verbunt2017} (within the 68~per cent confidence interval), the low-velocity component is inconsistent. We notice, that the sample compiled by \cite{Deller2019} strongly lacks low-velocity pulsars especially with large spin-down ages.
While the sample of \cite{bbgt02} contained two out of nine pulsars (roughtly 22~per~cent) with nominal transverse velocity less than 40~km~s$^{-1}$, our new sample has three out of 69 (4~per~cent) within the same velocity range. The reason for this is unclear. The low-velocity pulsars should stay close to the Galactic plane and, therefore, be abundant in any radio survey. One possible explanation is that pulsars in the sample by \cite{verbunt2017} and in the sample by \cite{Deller2019} have different spin-down ages. In Figure~\ref{f:tau}, we plot the spin-down ages distribution for these two samples. Overall, the \cite{Deller2019} sample contains fewer objects with $\tau < 10$~Myr and more older objects.

To check if an addition of secondary Maxwellian is significant, we compute the likelihood difference $d\mathcal L = 10$ which is equivalent to $\Delta \chi^2 = 99.3$~per cent for two degrees of freedom corresponding to two added parameters. This likelihood difference is very significant, but the significance decreased slightly in comparison to one found by \cite{verbunt2017}. We suspect that this decrease is caused by dearth of low-velocity pulsars in \cite{Deller2019} sample in comparison to the previous measurements.

\begin{table*}
    \centering
    \begin{tabular}{cccccccccccccccc}
    \hline
    h & H &     & \multicolumn{2}{c}{Sample}  &  \multicolumn{3}{c}{single Maxwellian} &  \multicolumn{7}{c}{two Maxwellians} \\
    & &   &  & N    &  $\sigma$ & range & $d\mathcal L$ & $\sigma_1$ & range & $\sigma_2$ & range & $w$ & range &  $d\mathcal L$ \\
    \hline     
   0.33 & 1.7 & Isotropic models & A & 69  & 229 & 215-245 & $\equiv 0$  & 146 & 125-162 & 317 & 277-346 & 0.58 & 0.47-0.68 & 10  \\
   0.32 & 0.8 & Isotropic models & A & 69  & 229 & 215-245 & 34          & 152 & 129-165 & 324 & 282-356 &  0.62 & 0.51-0.71  & 42    \\
   0.33 & 1.7 & Mixed models & A & 17+52 & 225 & 210-246 & 34 & 128 & 110-150 & 298 & 270-326 & 0.42 & 0.27-0.59 &  43 \\
    \hline
    
   0.33 & 1.7 & Isotropic models    & Y & 21  & 298 & 266-336 & $\equiv 0$  & 55  & 40-80 & 334 & 293-384 & 0.19 & 0.10-0.30 & 6   \\
   0.18 & 0.7 & Isotropic models    & Y & 21  & 290 & 259-326 & 28          & 55  & 40-80 & 325 & 286-374 & 0.19 & 0.10-0.30 & 34  \\
   0.33 & 1.7 & Mixed models    & Y & 6+15     & 298 & 266-334 & 18       & 70  & 50-113     & 328 & 288-375    & 0.17 & 0.08-0.29 & 21  \\   
   0.33 & 1.7 & J0614+2229 (isotropic) & Y & 7+14 & 295 & 264-332 & 19    & 56  & 41-81      & 336 & 291-381 & 0.20 & 0.10-0.31 &  25  \\
    & & other mixed \\
    \hline
    \end{tabular}
    \caption{Results of maximum likelihood analysis for 69 pulsars in our master list (A) and for the 21 youngest pulsars ($\tau < 3$~Myr; Y). Parameters $h$ and $H$ are used for the Galactic distribution of pulsars $f_D(D)$. Range is the 68 per cent confidence interval.}
    \label{t:res}
\end{table*}

Further, we introduce a mixed model: 17 pulsars with nominal $|z| < 50$~pc and pulsars with the spin-down age $\tau > 50$~Myr are analysed using an isotropic distribution and all other are analysed using a semi-isotropic distribution. The calculations of the semi-isotropic model used in \cite{verbunt2017} have to be slightly corrected to deal with much smaller errors for proper motions in the sample by \cite{Deller2019}. Namely, the parameter $h$ responsible for size of the region for the numerical integration (see Appendix C in \citealt{verbunt2017}) is decreased to $h = 2\pi / 500$. 

For single mixed Maxwellian we obtain $\sigma = 225_{-15}^{+21}$, for the mixed bimodal Maxwellian distribution we get following parameters: $w=42_{-15}^{+17}$, $\sigma_1 = 128_{-18}^{+22}$~km~s$^{-1}$ and $\sigma_2 = 298\pm 28$~km~s$^{-1}$.
We notice that when the mixed model is used, we obtain significantly lower values of the likelihood ($d\mathcal L = 34$ for single Maxwellian and $d\mathcal L = 43$ for bimodal Maxwellian). It means that younger pulsars outside of the $|z| < 50$~pc region indeed have preferable orientation of the velocity vector pointing outside of the Galactic plane as it is expected. Application of the mixed model shifts $w$, $\sigma_1$ and $\sigma_2$ to slightly lower values. In the case of bimodal Maxwellian, the mixed model keeps correlations between $w$ and $\sigma_1$, see Figure~\ref{f:contour_A} (lower left panel). 
The mixed bimodal Maxwellian model suggested in \cite{verbunt2017} is outside of the 99~per cent confidence interval. Addition of the second Maxwellian is significant because the likelihood difference of $d\mathcal L = 9$ is equivalent to $\Delta\chi^2 \approx 98.9$ per cent for two degrees of freedom. Therefore, introduction of the mixed model does not change the result qualitatively, but slightly shifts velocities to smaller values.

\subsection{Natal kicks of radio pulsars}
Studying the natal kick distribution requires us to restrict the sample to the youngest radio pulsars, which are less affected by observational selection and deceleration in the Galactic gravitational potential.
On the one hand, it is better to choose as small cut-off for spin-down ages as possible. The fastest pulsars ($\approx 1000$~km~s$^{-1}$) could travel up to 1~kpc per 1~Myr. Therefore, such pulsars could escape most modern radio surveys which concentrates on objects close to the Galactic plane. On the other hand, choosing inappropriately small age cut-off we unnecessary decrease the sample size. With this in mind, we decrease the cut-off age to 3~Myr (cut-off used by \citealt{hobbs2005}) in comparison to 10 Myr used in \cite{verbunt2017}. Our Y sample contains 21 objects which is comparable in size to Y sample from \cite{verbunt2017}. 

We perform multiple tests using the population synthesis code and additional filtration of radio pulsars to check if the velocity estimate works reliably when the observational selection is present. These tests are summarised in Appendix~\ref{s:tests}. We find that the method estimates the parameters of the natal kick velocity distribution reliably even if the selection is present. Moreover, $d\mathcal L \geq 6$ means a confident detection of the bimodality. From our tests it become clear that $d\mathcal L >2$ is already significant at $>90$ per cent level. The maximum likelihood estimates found for synthetic catalogues are not biased and well within expected confidence intervals.

When we optimise the likelihood function using the single isotropic Maxwelian velocity distribution with  sample of young pulsars we find that $\sigma = 298_{-32}^{+38}$~km~s$^{-1}$. The values found by \cite{hobbs2005} and \cite{verbunt2017} (single Maxwellian $\sigma=277$~km~s$^{-1}$) are within our 68~per cent confidence interval. It is worth noting that this value is significantly higher than $\sigma=229_{-14}^{+16}$~km~s$^{-1}$ found for the entire sample. This effect is likely a combination of the observational selection mentioned above and physical deceleration of pulsars in the Galactic gravitational potential.

When we optimise the model containing bimodal Maxwellian distribution we find a result within the 68~per cent confidence interval of \cite{verbunt2017}. The first Maxwellian with $\sigma_1 = 55_{-15}^{+25}$~km~s$^{-1}$ contains $19_{-9}^{+11}$ per cent of all objects and the second Maxwellian with $\sigma_2 = 334_{-41}^{+50}$~km~s$^{-1}$ contains $81_{-11}^{+9}$ per cent of all pulsars. The presence of a second component is quite significant: the likelihood difference $d\mathcal L = 6$ corresponds to $\Delta \chi^2 = 95$ per cent for two degrees of freedom. 

It is also worth noting that correlation between $\sigma_1$ and $w$ is practically negligible, see Figure~\ref{f:contour_A} (upper right panel). All three contours cover practically the same area, which includes the maximum likelihood point. It indicates that the low and high-velocity pulsars probably belong to clearly distinguished distributions. 

\begin{figure}
    \centering
    \includegraphics[width=1.0\linewidth]{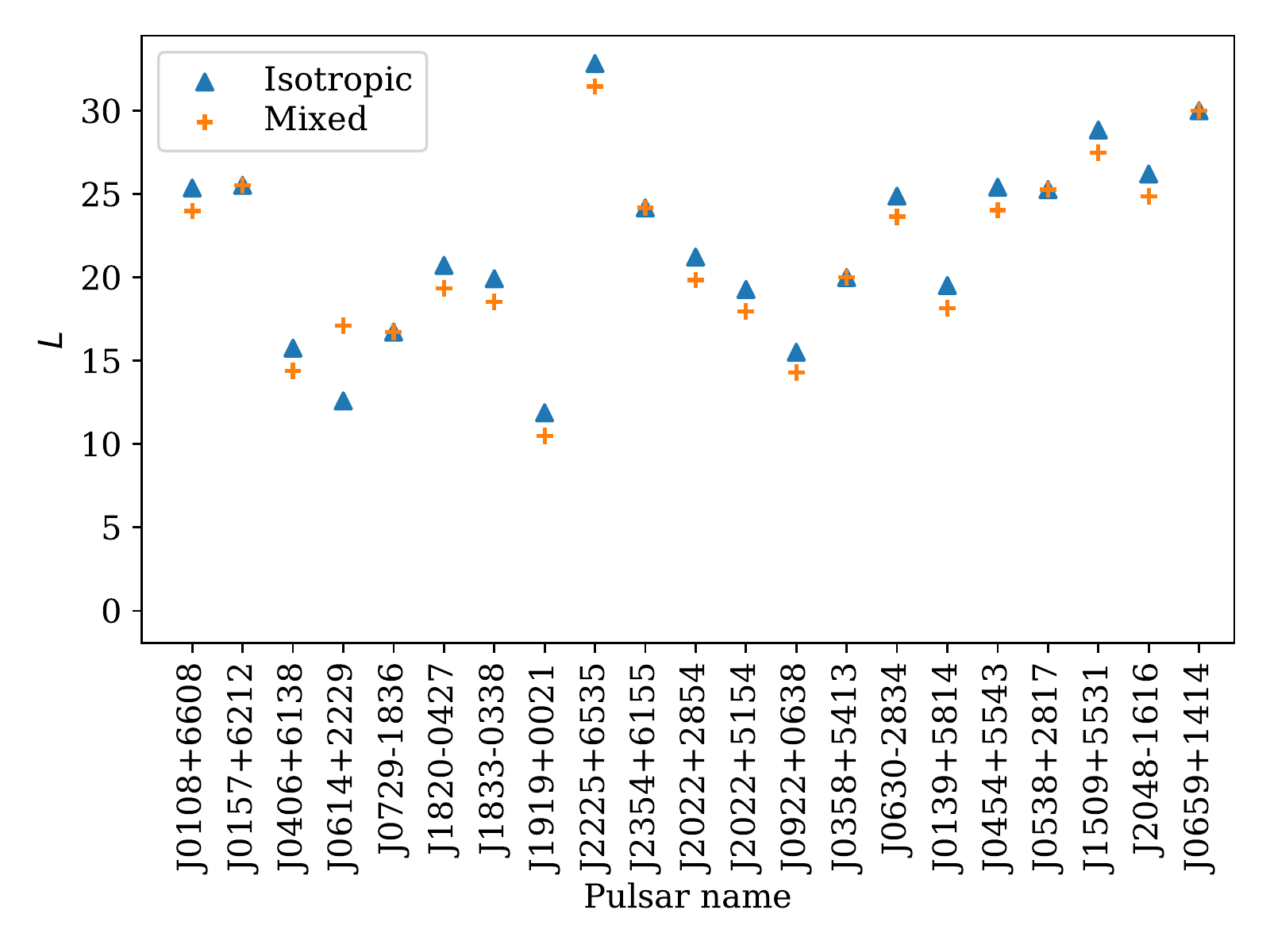}
    \caption{Comparison of individual contribution to the maximum likelihood for sample of radio pulsars with $\tau<3$~Myr. PSR J0614+2229 seems to strongly prefer isotropic model. To compute both  likelihoods we use parameters $w = 0.19$, $\sigma_1 = 55$ and $\sigma_2 = 334$.}
    \label{f:likelihood_compar}
\end{figure}

When we optimise the mixed model with single Maxwellian we obtain $\sigma = 298_{-32}^{+36}$ with $d\mathcal L = 18$ in comparison to single isotropic Maxwellian case. This model does not have any additional parameters. It means that the younger pulsars are clearly moving away from the Galactic plane. 

The bimodal mixed Maxwellian has following parameters: $w = 17_{-9}^{+12}$, $\sigma_1 = 70_{-20}^{+43}$~km~s$^{-1}$ and $\sigma_2 = 328_{-40}^{+47}$~km~$^{-1}$.
The parameters of bimodal mixed Maxwellian model are well within the $68$~per cent confidence interval of the isotropic model. The contours of constant likelihood are in Figure~\ref{f:contour_A} (right lower panel) resemble one found for the isotropic model with an exception of their much longer tails in the high-velocity direction. 

We were puzzled by the likelihood difference of this model $d\mathcal L = 21$ which is just $d\mathcal L = 3$ different from the single Maxwellian mixed model. Such a difference indicates only a slight preference ($\Delta \chi^2= 77.7$ per cent for two degrees of freedom\footnote{if we use assumption that the likelihood ratio test follows $\chi^2$ distribution precisely, but this result seems more significant if we take into account the results of Monte Carlo tests}). However, it is worth remembering that this model is based on at least two essential assumptions: (1) the velocity distribution is a sum of two Maxwellians and (2) the velocity vectors of young pulsars (beside ones in the $50$~pc stripe along the Galactic plane) are directed away from the plane. It seems that the second assumption plays an essential role in decreasing the likelihood difference. We plot the likelihood contributions of individual pulsars in Figure~\ref{f:likelihood_compar} and immediately notice that the mixed model containing two Maxwellians is in all cases better or comparable to two isotropic Maxwellians model except the case of PSR J0614+2229. 

PSR J0614+2229 has $\tau = 8.93\times 10^4$~years and $B = 4.5\times 10^{12}$~G estimated by magneto-dipole equation and nominal $z = \sin b / \varpi' = 148$~pc. Therefore, the PSR J0614+2229 should move away from the Galactic plane, while in reality it has a significant component of proper motion directed toward the plane. This is a reason why the significance of the mixed model with two Maxwellians is not high enough. Another object in the Y list which seems to move toward the plane is PSR J0538+2817. Both models have exactly the same likelihood for this pulsar which means that its proper motion toward the plane is apparent. This is not the case, however for J0614+2229.

\begin{figure}
    \centering
    \includegraphics[width=1.0\linewidth]{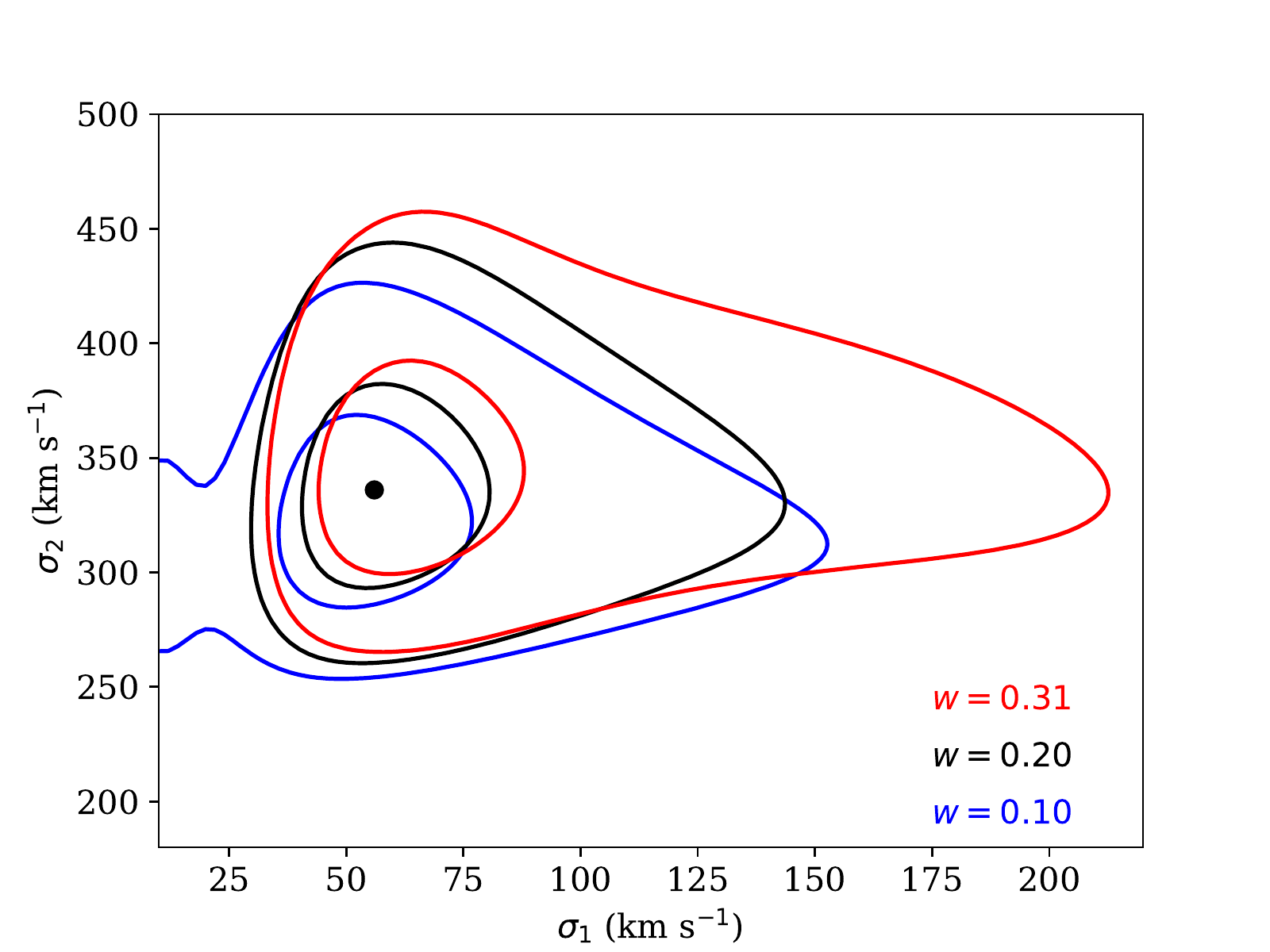}
    \caption{Contours of $d\mathcal L(\sigma)$ for Y sample and mixed model where PSR J0614+2229 is treated as one drawn from the isotropic velocity distribution.}
    \label{f:mmy_contour}
\end{figure}

We perform additional analysis moving PSR J0614+2229 into the isotropic list and 
obtain $w = 20_{-10}^{+11}$, $\sigma_1 = 56_{-15}^{+25}$~km~s$^{-1}$ and $\sigma_2 = 336\pm 45$~km~s$^{-1}$.
The results of likelihood optimisation in this case are well within 68 per cent confidence interval of the original mixed model. We plot the likelihood profile in Figure~\ref{f:mmy_contour}.
The plot is very similar to the isotropic case: long tails in the high-velocity direction seen in the original mixed model disappeared.
This model's total likelihood is significantly different. A mixed model with two Maxwellians with PSR J0614+2229 treated as a pulsar with velocity drawn from the isotropic velocity distribution is significantly more accurate than similar model which contains single Maxwellian ($d\mathcal L = 6$ corresponds to $\Delta \chi^2 = 95$ per cent for two degrees of freedom). 
Therefore, we conclude that PSR J0614+2229 has to be treated as an exception.

It is worth noting that the main reason our original mixed model with two Maxwellians is not as significant as new model is because the orientation of the velocity vector of J0614+2229 is unexpected.
If we completely remove J0614+2229 from the sample, we obtain same significance as in the case when we treat this pulsar as one with velocity drawn from the isotropic velocity distribution.

We run two additional optimisations for the case when $\varpi'/\sigma_\varpi > 2$ and $\tau < 3$~Myr, which adds one additional pulsar to the list. In the case of a single isotropic Maxwellian we get $\sigma=295_{-31}^{+37}$~km~s$^{-1}$. In the case of the sum of two isotropic Maxwellians we get $w = 0.17_{-9}^{+11}$, $\sigma_1 = 54_{-15}^{+26}$~km~s$^{-1}$ and $\sigma_2 = 327_{-39}^{+47}$~km~s$^{-1}$. The bimodal isotropic Maxwellians is more significant than a single Maxwellian with $d\mathcal{L} = 5$, which corresponds to 92 per cent probability. In comparison to our original result the shift in velocity estimates is about a few km~s$^{-1}$ and well within our confidence interval estimates.

\subsection{Galactic distribution of radio pulsars}
\begin{figure}
    \centering
    \includegraphics[width=1.0\linewidth]{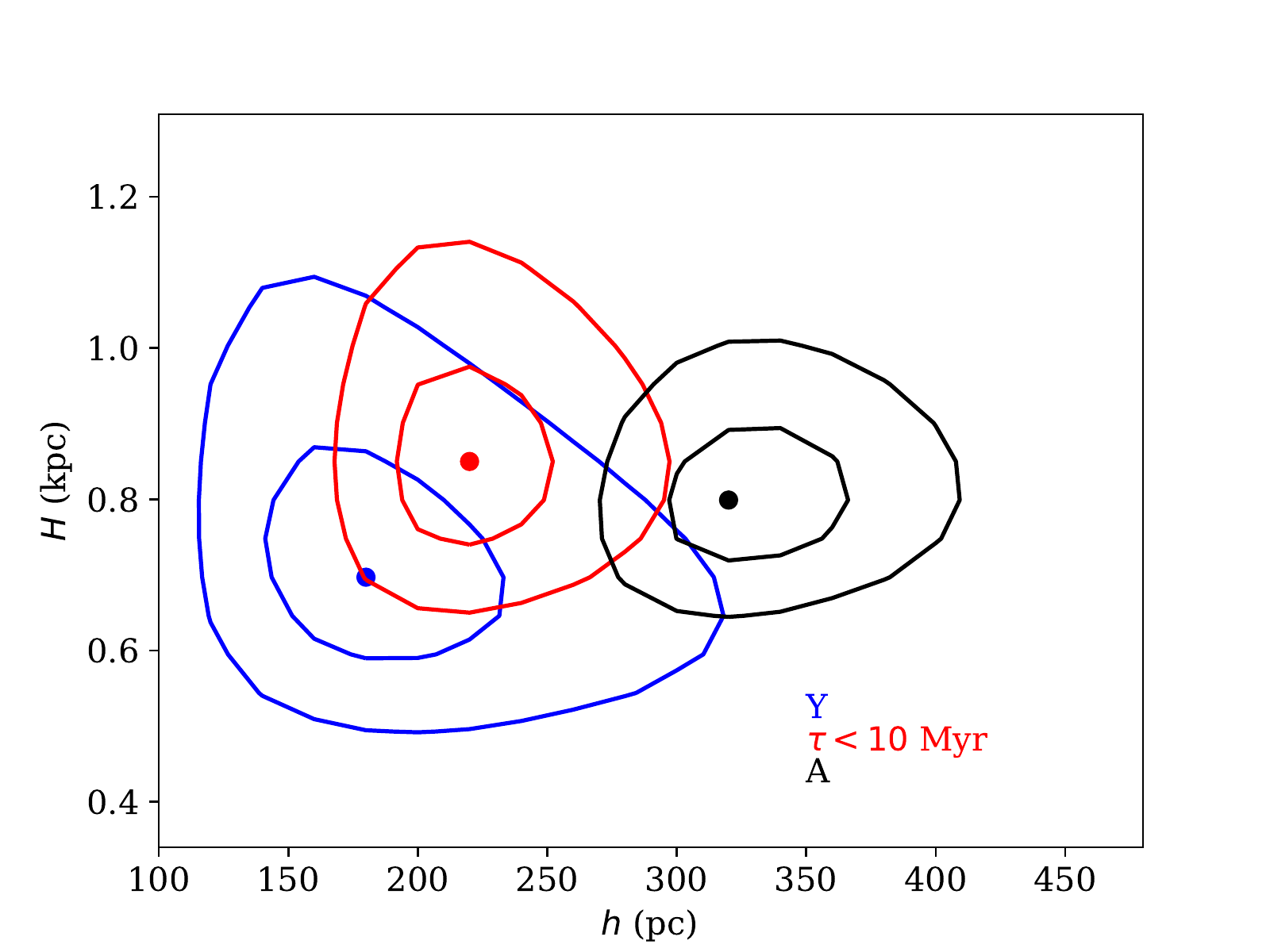}
    \caption{Contours of $d\mathcal L(\sigma)$ in the plane of $h$ and $H$ for A, I, Y samples}
    \label{f:hh_contour}
\end{figure}

We study the dependence of our maximum likelihood technique on parameters of the Galactic distribution of radio pulsars. To do so, we optimise the single isotropic Maxwellian model for values of $h$ and $H$ from eq.(\ref{e:fD}) computed on a grid. We perform this optimisation for A, Y sample and also for pulsars with $\tau<10$~Myr (sample I hereafter). Results are summarised in Table~\ref{t:hh} and in Figure~\ref{f:hh_contour}. 


\begin{table}
    \centering
    \begin{tabular}{ccccccc}
    \hline
    Sample  & $h$    & range     & $H$ & range \\
            & (kpc)  &           & (kpc) \\
    \hline        
    A       &  0.32  & 0.3-0.37  & 0.8  & 0.72-0.9  \\
    I       &  0.22  & 0.19-0.25 & 0.85 & 0.74-0.97 \\
    Y       &  0.18  & 0.15-0.23 & 0.70 & 0.59-0.87 \\
    \hline
    \end{tabular}
    \caption{Results of the maximum likelihood optimisation for parameters determining the Galactic distribution of radio pulsars for different samples. A includes all radio pulsars, I only ones with $\tau < 10$~Myr and Y only with $\tau < 3$~Myr. Range is 68 per cent confidence interval.}
    \label{t:hh}
\end{table}

We use the optimised values to estimate the parameters of the natal kick, see fifth line in Table~\ref{t:res}. Despite a dramatic change in value of parameters $h$ and $H$ the result of the maximum likelihood optimisation for velocity distribution is well within the 68 per cent confidence interval of original models, and parameters differ by less than 10~km~s$^{-1}$ (see Table~\ref{t:res}). This is related to the fact that parallaxes are measured with high precision in new PSR$\pi$ catalogue, therefore the initial distance distribution does not have a significant effect on velocities.

Our tests using the pulsar population synthesis code, see Apendix~\ref{s:tests} show that the values of $h$ and $H$ are very sensitive to exact distribution of latitudes of measured pulsars. In particular, if pulsars are too concentrated toward the Galactic plane, the method tends to give smaller $h$ values. On the other hand, even if the values of $h$ and $H$ values are biased, the method estimates parameters of the velocity distribution correctly.

\section{Discussion}

\begin{figure}
    \centering
    \includegraphics[width=1.0\linewidth]{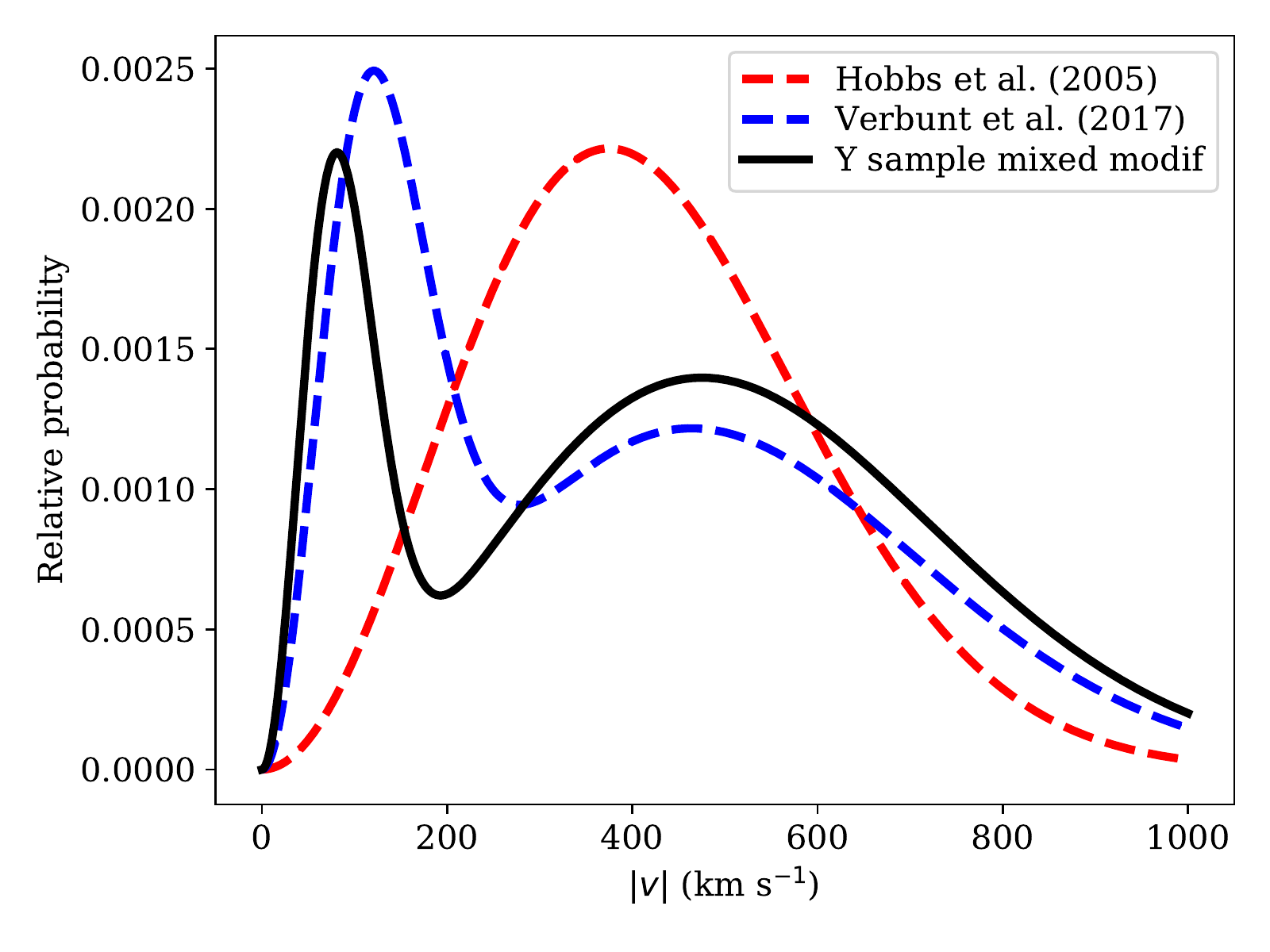}
    \caption{The probability density functions for natal kicks found in different studies. Black solid line shows the distribution found for young radio pulsars using the mixed model with an exception of J0614+2229 which is treated as drawn from the isotropic velocity distribution. Red dashed line shows the distribution found by \protect\cite{hobbs2005} and blue dashed lines shows one found by \protect\cite{verbunt2017}.}
    \label{f:distr_three}
\end{figure}

\begin{figure}
    \centering
    \includegraphics[width=1.0\linewidth]{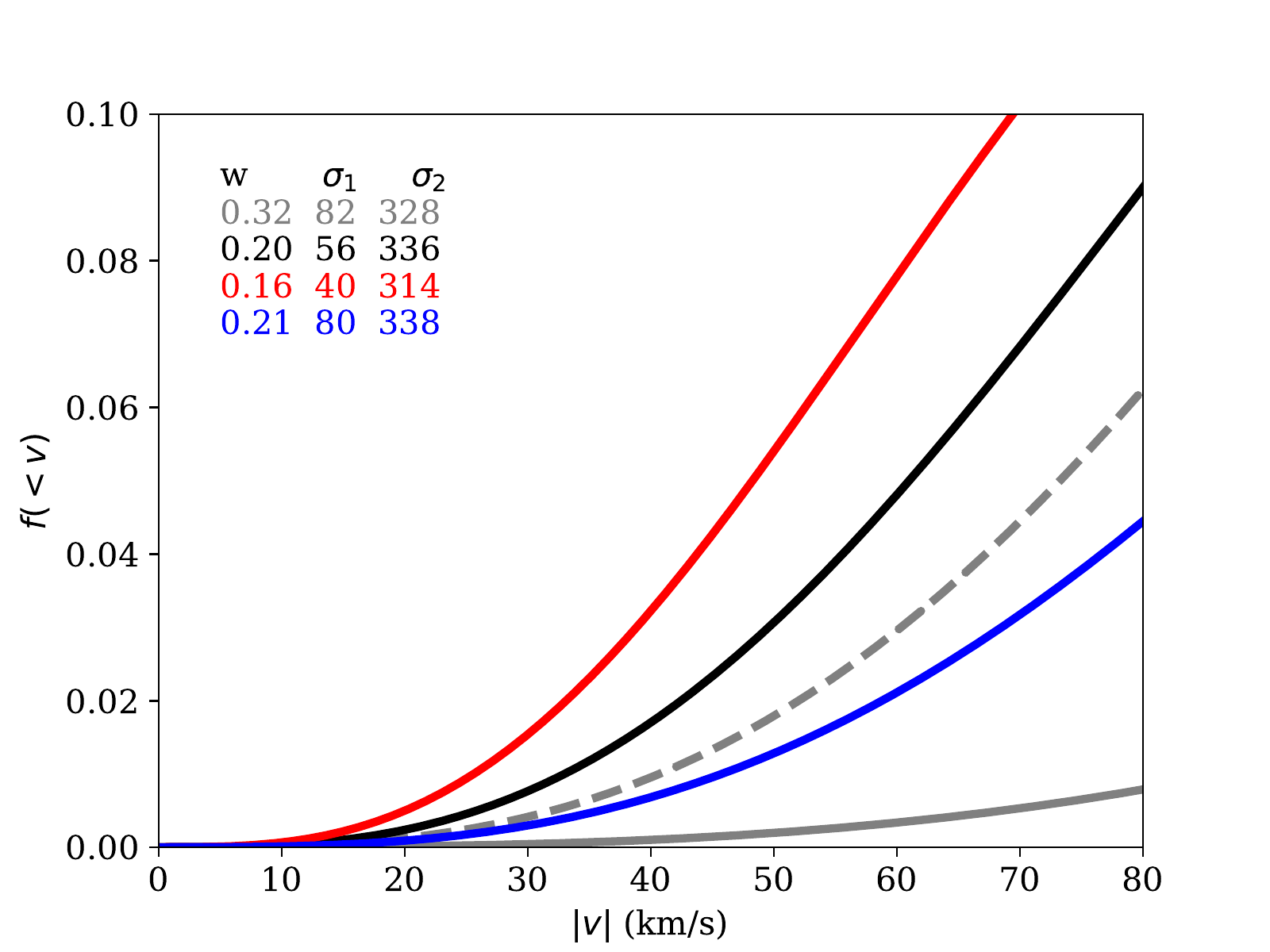}
    \caption{Fraction of young radio pulsars with velocities less than $|v|$. Solid black line shows the best mixed model for $Y$ sample and red and blue solid lines show the best models for variations of $\sigma_1$. The grey dashed line shows results from \protect\cite{verbunt2017} and dashed solid line is for distribution by \protect\cite{hobbs2005} with $\sigma=265$~km~s$^{-1}$.}
    \label{f:fract}
\end{figure}

In Figure~\ref{f:distr_three} we plot the result of our analysis and two works mentioned previously. Our result differs significantly from the \cite{hobbs2005} and more similar to work by \cite{verbunt2017}. In comparison to this latter research, the first peak of the velocity distribution has shifted slightly toward even smaller velocities, the height of the first peak decreased and the gap between two modes becomes more apparent. Using our best model for velocities of young radio pulsars, we plot the fraction of NSs born with $v<60$~km~s$^{-1}$ in Figure~\ref{f:fract}. This fraction varies between 2 and 8 per cent and is compatible with one found before in \cite{verbunt2017}.

It is important to highlight once again that our analysis is based only on parallaxes and proper motion measured for \textit{isolated} radio pulsars \footnote{see for example \cite{igoshev2019} for search for ultra-wide companions for these objects}. If a weak natal kick (0-60~km~s$^{-1}$) occurs more frequently (or solely) in interacting binaries, the binary often stays bound and, therefore, is excluded from our master list. There are a number of theoretical mechanisms suggested in support of this scenario, such as the electron capture supernova explosion or accretion induced WD collapse. The kick amplitude might be related to an amount of mass lost during the explosion, so a stripped star in an interacting binary would receive much weaker kick and ends up as an isolated pulsar only in exceptional cases.
With further constraints on natal kick distributions derived for merging double NS systems (see e.g. \citealt{lowkickNSNS}), MSPs and BeX binaries it is important to check if the tension between results of our method and those measurements keeps growing. If it is the case, the preferable binary origin of weak natal kicks seems to be the most plausible explanation.

There is an additional caveat in our analysis. The Galactic gravitational potential affects velocity of isolated neutron stars. Stars moving at the highest speeds (>500 km~s$^{-1}$) leave the Galaxy or become a part of the Galactic halo. On the other hand, the only quantity which is available for us to distinguish between older and younger pulsars is the spin-down age. This age is affected by the magnetic field evolution and initial periods. If we assume that our understanding of these two quantities is reasonable, we do not expect the spin-down age to be extremely different from the actual age (for example, there cannot be a young pulsar with $\tau = 1$~Gyr). 
However, if there is an unknown mechanism which mixes spin-down ages e.g the magntic field re-emergence after a fall-back \citep{igoshev2016}, our selection of youngest pulsars can be more strongly affected by the observational selection.

It is worth to discuss the case of PSR J0614+2229. There are two possible explanations for its abnormal velocity vector directed toward the Galactic plane: (1) pulsar was born high above the plane or (2) pulsar experienced unusual magnetic field evolution. In the first case the pulsar might have been a part of a massive binary system which received large systemic velocity (order of $100$~km~s$^{-1}$) during the first supernova explosion. In this case the binary had some time to travel away from the Galactic plane before the secondary supernova explosion.  The secondary explosion producing the observed pulsar  disrupted the binary. In the second case, PSR J0614+2229 might be a pulsar with a hidden magnetic field which has re-merged over time which makes the simple spin-down age estimate extremely unreliable.

\section{Conclusion}

We analysed new parallaxes and proper motion measurements for isolated radio pulsars published by \cite{Deller2019} together with older measurements using the maximum likelihood method.
We optimised the parameters of velocity distribution of pulsars.
We find:
\begin{description}
    \item The velocity distribution of pulsars of all ages presented in our master list is compatible with a bimodal Maxwellian with contribution of first mode $w = 0.42_{-0.15}^{+0.17}$, $\sigma_1 = 128_{-18}^{+22}$~km~s$^{-1}$ and second mode $\sigma_2 = 298\pm 28$~km~s$^{-1}$. It differs from results of \cite{verbunt2017} and might be explained by a different selection of pulsars in \cite{Deller2019} in comparison to earlier works. The presence of bimodality is significant at the level of 99.3 per cent.
    \item The natal kick distribution of NSs derived as the velocity distribution of young radio pulsars ($\tau < 3$~Myr) can be described as bimodal Maxwellian with $w = 0.2_{-0.10}^{+0.11}$, $\sigma_1=56_{-15}^{+25}$~km~s$^{-1}$ and $\sigma_2 = 336\pm 45$~km~s$^{-1}$. This distribution lies well within the 68 per cent confidence interval of the distribution found by \cite{verbunt2017}. The presence of bimodality is significant at the level of 95 per cent. 
    \item We perform multiple Monte Carlo simulations using a code for the pulsar population synthesis and show that the observational selection does not affect the estimated parameters of the natal kick distribution.
    \item The results of the maximum likelihood method for parameters of velocity distribution are weakly sensitive to the exact value of $h$ and $H$ used and changes are within 10~km~s$^{-1}$ if alternative values are assumed.
    \item The fraction of radio pulsars born with $|v| < 60$~km~$^{-1}$ is $5\pm 3$ per cent in our best model.
    \item We notice that PSR J0614+2229 with spin-down age $\tau \approx 9\times 10^4$ years moves toward the Galactic plane from its nominal height of $z \approx 150$~pc. It could indicate a complicated binary origin of this radio pulsar or its non-standard magnetic field evolution.
\end{description}

\section*{Acknowledgements}
A.P.I. acknowledges the guidance and help he received from Prof. Eric Cator and Prof. Frank Verbunt while working on codes used in this research. A.P.I. thanks the Science and Technology Facilities Council grant ST/S000275/1. A.P.I. is grateful to Prof. S.B. Popov and A. Frantsuzova for fruitful discussions about this manuscript. A.P.I. thanks anonymous referee for multiple comments, which helped to improve the manuscript significantly.

\bibliographystyle{mnras}
\bibliography{ref_vel_psr2} 




\appendix

\section{List of pulsars}

\begin{table*}
\begin{tabular}{llcrrccrrrc}
\hline
B-name & J-name & & $l$ & $b$ & $P$ & $\log \tau$ & $\mu_{\alpha*,o}\, \sigma_{\alpha}$ & $\mu_{\delta, o}\, \sigma_{\delta}$  &  $\varpi \, \sigma_\varpi$ & Ref.\\
&     &  & ($^\circ$) &  ($^\circ$) & (s) &  (yr)         & (mas/yr)                   & (mas/yr)                    &  (mas)    \\
\hline
B0031-07 & J0034-0721  & &  110.42  &  -69.82  &  0.94295  &  7.56  &  10.37(8)  &  -11.13(16)  &  0.93(8)  &  6  \\
B0052+51 & J0055+5117  & &  123.62  &  -11.58  &  2.11517  &  6.55  &  10.490(85)  &  -17.352(204)  &  0.349(55)  &  8  \\
B0059+65 & J0102+6537  & &  124.08  &  2.77  &  1.67916  &  6.65  &  9.252(81)  &  1.828(206)  &  0.399(45)  &  8  \\
B0105+65 & J0108+6608  & &  124.65  &  3.33  &  1.28366  &  6.19  &  -32.754(36)  &  35.162(51)  &  0.468(35)  &  8  \\
B0136+57 & J0139+5814  & &  129.22  &  -4.04  &  0.27245  &  5.61  &  -19.11(7)  &  -16.60(7)  &  0.37(4)  &  6  \\
\\
B0144+59 & J0147+5922  & &  130.06  &  -2.72  &  0.19632  &  7.08  &  -6.380(101)  &  3.826(97)  &  0.495(93)  &  8  \\
B0154+61 & J0157+6212  & i &  130.59  &  0.33  &  2.35174  &  5.3  &  1.521(105)  &  44.811(48)  &  0.554(39)  &  8  \\
B0320+39 & J0323+3944  & i &  152.18  &  -14.34  &  3.03207  &  7.88  &  26.484(59)  &  -30.780(29)  &  1.051(40)  &  8  \\
B0329+54 & J0332+5434  & i &   145.0  &  -1.22  &  0.71452  &  6.74  &  16.969(29)  &  -10.379(58)  &  0.595(25)  &  8  \\
B0331+45 & J0335+4555  & i &  150.35  &  -8.04  &  0.2692  &  8.76  &  -3.638(73)  &  -0.097(134)  &  0.409(27)  &  8  \\
\\
B0353+52 & J0357+5236  & i &  149.1  &  -0.52  &  0.19703  &  6.82  &  13.908(115)  &  -10.633(98)  &  0.305(77)  &  8  \\
B0355+54 & J0358+5413  & i &  148.19 &  0.81  &  0.15638  &  5.75  &  9.20(18)  &  8.17(39)  &  0.91(16)  &  4  \\
B0402+61 & J0406+6138  & &  144.02  &  7.05  &  0.59458  &  6.23  &  12.400(151)  &  22.716(100)  &  0.218(57)  &  8  \\
B0450+55 & J0454+5543  & &  152.62  &  7.55  &  0.34073  &  6.36  &  53.34(6)  &  -17.56(14)  &  0.84(5)  &  6  \\
         & J0538+2817  & i &  179.72  &  -1.69  &  0.14316  &  5.79  &  -23.57(10)  &  52.87(10)  &  0.72(12)  &  6  \\
\\
B0559-05 & J0601-0527  & &  212.2  &  -13.48  &  0.39597  &  6.68  &  -7.348(77)  &  -15.227(105)  &  0.478(45)  &  8  \\
B0611+22 & J0614+2229  & i* &  188.79  &  2.4  &  0.33496  &  4.95  &  -0.233(53)  &  -1.224(65)  &  0.282(31)  &  8  \\
B0626+24 & J0629+2415  & &  188.82  &  6.22  &  0.47662  &  6.58  &  3.629(193)  &  -4.607(153)  &  0.333(54)  &  8  \\
B0628-28 & J0630-2834  & &  236.95  &  -16.76  &  1.24442  &  6.44  &  -46.30(99)  &  21.26(52)  &  3.009(409)  &  5  \\
B0656+14 & J0659+1414  & i &  201.11  &  8.26  &  0.38489  &  5.05  &  44.07(63)  &  -2.40(29)  &  3.47(36)  &  2  \\
\\
B0727-18 & J0729-1836  & i &  233.76  &  -0.34  &  0.51016  &  5.63  &  -13.072(125)  &  13.252(456)  &  0.489(98)  &  8  \\
B0809+74 & J0814+7429  & i & 140.0  &  31.62  &  1.29224  &  8.09  &  24.02(9)  &  -43.96(35)  &  2.31(4)  &  1  \\
B0818-13 & J0820-1350  & &  235.89  &  12.59  &  1.23813  &  6.97  &  21.64(9)  &  -39.44(5)  &  0.51(4)  &  6  \\
B0823+26 & J0826+2637  & &  196.96  &  31.74  &  0.53066  &  6.69  &  62.994(21)  &  -96.733(85)  &  2.010(13)  &  8  \\
B0919+06 & J0922+0638  & & 225.42  &  36.39  &  0.43063  &  5.7  &  18.8(9)  &  86.40(70)  &  0.83(13)  &  3  \\
\\
B0950+08 & J0953+0755  & &  228.91  &  43.7  &  0.25307  &  7.24  &  -2.09(8)  &  29.46(7)  &  3.82(7)  &  1  \\
B1133+16 & J1136+1551  & &  241.9  &  69.2  &  1.18791  &  6.7  &  -73.785(31)  &  366.569(72)  &  2.687(18)  &  8  \\
B1237+25 & J1239+2453  & &  252.45  &  86.54  &  1.38245  &  7.36  &  -106.82(17)  &  49.92(18)  &  1.16(8)  &  1  \\
B1322+83 & J1321+8323  & &  121.89  &  33.67  &  0.67004  &  7.27  &  -52.674(99)  &  32.373(204)  &  0.968(140)  &  8  \\
B1508+55 & J1509+5531  & & 91.33  &  52.29  &  0.73968  &  6.37  &  -73.64(5)  &  -62.65(9)  &  0.47(3)  &  6  \\
\\
B1530+27 & J1532+2745  & &  43.48  &  54.5  &  1.12484  &  7.36  &  1.542(127)  &  18.932(118)  &  0.624(96)  &  8  \\
B1541+09 & J1543+0929  & &  17.81  &  45.78  &  0.74845  &  7.44  &  -7.61(6)  &  -2.87(7)  &  0.13(2)  &  6  \\
B1540-06 & J1543-0620  & &  0.57  &  36.61  &  0.70906  &  7.11  &  -16.774(63)  &  -0.312(147)  &  0.322(45)  &  8  \\
B1556-44 & J1559-4438  & &  334.54  &  6.37  &  0.25706  &  6.6  &  1.52(14)  &  13.15(5)  &  0.384(81)  &  5  \\
B1604-00 & J1607-0032  & &  10.72  &  35.47  &  0.42182  &  7.34  &  -26.437(99)  &  -27.505(222)  &  0.934(47)  &  8  \\
\\
B1620-09 & J1623-0908  & & 5.3  &  27.18  &  1.27644  &  6.89  &  -10.769(131)  &  23.509(166)  &  0.586(101)  &  8  \\
B1642-03 & J1645-0317  & &  14.11  &  26.06  &  0.38769  &  6.54  &  -1.011(51)  &  20.523(205)  &  0.252(28)  &  8  \\
B1700-18 & J1703-1846  & &  3.23  &  13.56  &  0.80434  &  6.87  &  -0.751(102)  &  16.962(230)  &  0.348(49)  &  8  \\
B1732-07 & J1735-0724  & &  17.27  &  13.28  &  0.41933  &  6.74  &  0.791(87)  &  20.614(74)  &  0.150(41)  &  8  \\
B1738-08 & J1741-0840  & &  16.96  &  11.3  &  2.04308  &  7.15  &  0.436(126)  &  6.876(109)  &  0.279(58)  &  8  \\
\\
B1753+52 & J1754+5201  & &  79.61  &  29.63  &  2.3914  &  7.38  &  -3.950(47)  &  1.101(72)  &  0.160(29)  &  8  \\
B1818-04 & J1820-0427  & &  25.46  &  4.73  &  0.59808  &  6.18  &  -7.318(74)  &  15.883(88)  &  0.351(55)  &  8  \\
B1831-03 & J1833-0338  & &  27.66  &  2.27  &  0.6867  &  5.42  &  -17.409(158)  &  15.038(337)  &  0.408(67)  &  8  \\
B1839+56 & J1840+5640  & &  86.08  &  23.82  &  1.65286  &  7.24  &  -31.212(33)  &  -29.079(82)  &  0.657(65)  &  8  \\
         & J1901-0906  & &  25.98  &  -6.44  &  1.78193  &  7.24  &  -7.531(45)  &  -18.211(159)  &  0.510(67)  &  8  \\
\\
B1911+13 & J1913+1400  & &  47.88  &  1.59  &  0.52147  &  7.01  &  -5.265(72)  &  -8.927(65)  &  0.185(27)  &  8  \\
B1917+00 & J1919+0021  & &  36.51  &  -6.15  &  1.27226  &  6.42  &  10.167(143)  &  -4.713(102)  &  0.166(42)  &  8  \\
B1929+10 & J1932+1059  & i &  47.38  &  -3.88  &  0.22652  &  6.49  &  94.06(9)  &  43.24(17)  &  2.78(6)  &  7  \\
B1935+25 & J1937+2544  & &  60.84  &  2.27  &  0.20098  &  6.7  &  -10.049(42)  &  -13.055(39)  &  0.318(31)  &  8  \\
B2003-08 & J2006-0807  & i &  34.1  &  -20.3  &  0.58087  &  8.3  &  -6.176(70)  &  -10.616(174)  &  0.424(101)  &  8  \\
\hline
\end{tabular}
\label{t:master}
\caption{Ref. is the number from the Table~\ref{t:lit_sourc}. An i in second column indicates that an isotropic velocity  distribution was used to model this particular pulsar in the mixed velocity model. For PSR J0614+2229 the isotropic model was used only in one case, see Section 4.2 for details.}
\end{table*}

\begin{table*}
\begin{tabular}{llcrrccrrrc}
\hline
B-name & J-name & & $l$ & $b$ & $P$ & $\log \tau$ & $\mu_{\alpha*,o}\, \sigma_{\alpha}$ & $\mu_{\delta, o}\, \sigma_{\delta}$  &  $\varpi \, \sigma_\varpi$ & Ref. \\
&      & & ($^\circ$)  &  ($^\circ$) & (s) &  (yr)         & (mas/yr)                   & (mas/yr)                    &  (mas)    \\
\hline
B2016+28 & J2018+2839  & i &  68.1  &  -3.98  &  0.55795  &  7.78  &  -2.64(21)  &  -6.17(38)  &  1.03(10)  &  1  \\
B2020+28 & J2022+2854  & &  68.86  &  -4.67  &  0.3434  &  6.46  &  -3.46(17)  &  -23.73(21)  &  0.61(8)  &  7  \\
B2021+51 & J2022+5154  & &  87.86  &  8.38  &  0.5292  &  6.44  &  -5.03(27)  &  10.96(17)  &  0.78(7)  &  7  \\
B2044+15 & J2046+1540  & i &  61.11  &  -16.84  &  1.13828  &  8.0  &  -10.455(90)  &  0.681(90)  &  0.310(82)  &  8  \\
B2043-04 & J2046-0421  & &  42.68  &  -27.39  &  1.54694  &  7.22  &  10.760(38)  &  -4.404(373)  &  0.167(42)  &  8  \\
\\
B2045-16 & J2048-1616  & &  30.51  &  -33.08  &  1.96157  &  6.45  &  113.16(2)  &  -4.60(28)  &  1.05(3)  &  6  \\
B2053+36 & J2055+3630  & &  79.13  &  -5.59  &  0.22151  &  6.98  &  1.04(4)  &  -2.46(13)  &  0.17(3)  &  6  \\
B2110+27 & J2113+2754  & i &  74.99  &  -14.03  &  1.20285  &  6.86  &  -27.981(52)  &  -54.432(96)  &  0.704(23)  &  8  \\
B2111+46 & J2113+4644  & &  89.0  &  -1.27  &  1.01468  &  7.35  &  9.525(148)  &  8.846(90)  &  0.454(77)  &  8  \\
         & J2144-3933  & i &  2.79  &  -49.47  &  8.50983  &  8.43  &  -57.89(88)  &  -155.90(54)  &  6.051(560)  &  5  \\
\\
B2148+63 & J2149+6329  & &  104.25  &  7.41  &  0.38014  &  7.55  &  15.786(131)  &  11.255(284)  &  0.356(72)  &  8  \\
B2154+40 & J2157+4017  & &  90.49  &  -11.34  &  1.52527  &  6.85  &  16.13(10)  &  4.12(12)  &  0.28(6)  &  6  \\
B2224+65 & J2225+6535  & &  108.64  &  6.85  &  0.68254  &  6.05  &  147.220(243)  &  126.532(115)  &  1.203(204)  &  8  \\
         & J2248-0101  & &  69.26  &  -50.62  &  0.47723  &  7.06  &  -10.548(117)  &  -17.407(267)  &  0.256(67)  &  8  \\
B2303+30 & J2305+3100  & &  97.72  &  -26.66  &  1.57589  &  6.94  &  -3.737(82)  &  -15.571(163)  &  0.223(33)  &  8  \\
\\
B2310+42 & J2313+4253  & &  104.41  &  -16.42  &  0.34943  &  7.69  &  24.15(10)  &  5.95(13)  &  0.93(7)  &  6  \\
B2315+21 & J2317+2149  & &  95.83  &  -36.07  &  1.44465  &  7.34  &  8.522(104)  &  0.136(192)  &  0.510(57)  &  8  \\
         & J2346-0609  & &  83.8  &  -64.01  &  1.18146  &  7.14  &  37.390(42)  &  -20.230(107)  &  0.275(36)  &  8  \\
B2351+61 & J2354+6155  & i &  116.24  &  -0.19  &  0.94478  &  5.96  &  22.755(56)  &  4.888(33)  &  0.412(43)  &  8  \\
\hline
\end{tabular}
\label{t:master1}
\caption{Ref. is the number from the Table~\ref{t:lit_sourc}. An i in second column indicates that an isotropic velocity  distribution was used to model this particular pulsar in the mixed velocity model.}
\end{table*}

\section{Testing the method}
\label{s:tests}

Some initial tests of the maximum likelihood methods are described in Chapter 3 of \cite{igoshev2017_phd}. These tests included limited or absent observational selection.
Here we perform new tests taking into account observation selection typical for isolated radio pulsars. For each test we perform following procedure: (1) we generate ten synthetic catalogues (each containing 21 objects) with fixed velocity distribution and errors similar to ones in our catalogue, (2) we run code for single isotropic Maxwellian model, (3) we run code for bimodal isotropic Maxwellian model and (4) we compare results of optimisation with nominal values. The main question which we try to answer with these tests if the results of the method are indicative of the actual natal kick distribution.

A synthetic catalogue is prepared in following manner: (1)  run the population synthesis code for isolated radio pulsars, (2)  filter the resulting catalogue selecting close radio pulsars with spin-down ages less than 3 Myr, (3)  convert the catalogue in the format accepted by our maximum likelihood code and add random value to  parallax and proper motions.

To obtain synthetic pulsars we use the isolated pulsar population synthesis code \texttt{NINA}\footnote{Source code and instruction is available https://github.com/ignotur/NINA} (Nova Investigii Neutronicorum Astrorum - New Studies of Neutron Stars). Some previous usage and description of this code can be found in \cite{igoshev2011,igoshev_gmm,igoshevPeriods}. The code essentially follows the algorithm described by \cite{faucher2006}. There are small changes, which we summarise here. The code traces isolated stellar evolution using approximated equations by \cite{Hurley2000}. Massive stars are born in spiral arms on a circular orbit around the Galactic center. Additional velocity is drawn from the Maxwellian velocity distribution with $\sigma = 15$~km~s$^{-1}$ to mimic the spread in OB stars velocities.  
The stellar evolution and motion in the Galactic gravitational potential \citep{Kuijken1989} is traced up to the supernova explosion. At the moment of supernova explosion a natal kick is imparted to the NS. The velocity vector of the kick is added to the current velocity in the Galactic gravitational potential. The motion of produced NS is integrated further. We populate the Galaxy with NS starting 300 Myr ago. To compute the period and period derivative, we assume that the initial magnetic field is drawn from the log-normal distribution with $\mu_B = 12.65$ [$\log_{10}$~G] and $\sigma_B = 0.55$ while the initial periods are drawn from the normal distribution with $\mu_p = 0.25$~s and $\sigma_p = 0.15$. We assume that the magnetic field is constant and does not decay. We use all parameters for the radio luminosity as they found by \cite{faucher2006}. We perform the same selection as in aforementioned article to identify radio bright pulsars which could have been detected in the the Parkes Multibeam Survey or Swinburn Multibeam survey \citep{parkesMultibeam, SwinburnMultibeam}.

These synthetic pulsars are located mostly at much larger distances than pulsars with measured parallaxes and proper motions. Therefore, we filter this catalogue selecting a subsample of pulsars, which approximately follows the distance distribution seen in our sample. Namely, we prepare the cumulative distribution of measured parallaxes for stars in our sample. Further for each synthetic pulsar we generate a uniform random value between 0 and 1. If this value is less than the cumulative fraction measured in our sample for the parallax of the synthetic pulsar, we accept the synthetic pulsar in our sample. Otherwise, we disregard it.

At this step we compute galactic coordinates of synthetic pulsars and their proper motions.
For consistency, we add a normal distributed variable to actual parallax and proper motion of the pulsar with the standard deviations equal to errors in the real catalogue.

We perform tests for both false positive and false negative rates of our method. The false positive rate is how often the method wrongly claims that the natal kick velocity distribution is bimodal while the actual distribution contains a single Maxwellian. The false negative rate is how often the method claims that the velocity distribution is single Maxwellian while the actual distribution is bimodal.

To study the false positive cases we generate ten radio pulsar populations with assumed single Maxwellian velocity distribution with $\sigma=256$~km~s$^{-1}$. For all ten cases, the single Maxwellian velocity distribution is the preferred velocity distribution. In nine cases we obtain $d\mathcal{L} = 0$ for the sum of two Maxwellians, which means that addition of two parameters do not improve the fit at all. In a single case we obtained $d \mathcal L = 1.32$ which does not show any significant improvement in the fit. The estimated values for sigma in all cases were within 120~km~s$^{-1}$ confidence interval centred at the assumed value. In 6 out of 10 cases the estimated $\sigma$ was within 60~km~s$^{-1}$ confidence interval. Based on this test we can conclude that the false positive rate of the method is very small ($<10\%$ for $d \mathcal L > 1.32$), so the method will not discover bimodal Maxwellian instead of a single Maxwellian. In the real sample we identified two Maxwellians with $d \mathcal L = 6$.

To test the false negative rate of the method we performed two tests varying the velocity distribution. In the first test we assumed that the natal kick distribution consists of two Maxwellians with a half of objects drawn from a Maxwellain with $\sigma_1 = 40$~km~s$^{-1}$ while other half is drawn from a Maxwellain with $\sigma_2 = 300$~km~s$^{-1}$. 
We analyse the synthetic catalogue with our code. The sum of two Maxwellians always gives a closer fit to the distribution.
Only in one case we identify $d \mathcal L = 3.2$, so it is somewhat small to be absolutely confident in the result. In all other cases $d\mathcal L = 12-28$. In the result the estimated fraction of low velocity objects ranges from 15 to 62 per cent. Therefore, under these conditions the false negative rate is very small (<10\% for $d\mathcal L > 3.2$).

We test our method for the velocity distribution that was identified as the best model for the natal kick, i.e. we assume that 19 per cent of radio pulsars are drawn from the Maxwellian with $\sigma_1 = 55$~km~s$^{-1}$ and remaining 81 per cent is from Maxwellian with $\sigma_2 = 334$~km~s$^{-1}$. In two cases we find that the Maxwellian velocity distribution describes the data from the synthetic catalogue well enough $d\mathcal L = 0$. In total of 6 cases $d\mathcal L < 5$, so if we did not know the underlying velocity distribution, we would suggest that this is a simple Maxwellian. Therefore, for the realistic natal kick parameters the false negative rate is quite large ($\approx 60$ per cent for $d \mathcal L < 5$). Taking into account all results, we can conclude that it is very improbable to find two Maxwellians for the natal kick velocity distribution in the data if only one is present, but it is somewhat probable to find a single Maxwellian in the data even if two Maxwellians are actually present and the contribution of weak natal kicks is small. Overall, severe selection effects imposed on the pulsar population do not seem to affect the result of our analysis for found values of $d\mathcal L$.

To test if the method correctly estimates $h$ and $H$ values we proceed in two steps. First of all, we generate a synthetic catalogue following all assumptions of the maximum likelihood technique, so no observational selection. In particular we distributed pulsars isotropically on the sky. In this case the maximum likelihood technique correctly estimated $h$ and $H$ values. Secondly, we use synthetic catalogues from previous test. In these tests, we find that the estimate for $h$ is about of 50-100~pc while in reality the assumptions of the population synthesis give $h_\mathrm{true}\approx 300$~pc. The reason for this mismatch is that the pulsars are not distributed uniformly over the Galactic latitude in the population synthesis, but instead follow band with $|b| < 15^\circ$. We try to modify the method using the integrated likelihood approach \citep{integLik} for parameters of the velocity distribution and for Galactic latitudes. Both options did not work in our tests. Therefore, we cannot prove that the estimated $h$ and $H$ are indeed correct for our sample.
 

\bsp	
\label{lastpage}
\end{document}